\documentclass[]{interact}

\usepackage{epstopdf}% To incorporate .eps illustrations using PDFLaTeX, etc.
\usepackage{subfigure}% Support for small, `sub' figures and tables

\usepackage{natbib}% Citation support using natbib.sty
\bibpunct[, ]{(}{)}{;}{a}{}{,}% Citation support using natbib.sty
\renewcommand\bibfont{\fontsize{10}{12}\selectfont}% Bibliography support using natbib.sty

\usepackage{hyperref}
\usepackage[ruled,vlined]{algorithm2e}

\SetCommentSty{mycommfont}
\usepackage{amsmath}
\usepackage{amsfonts}
\usepackage{amsthm}
\usepackage{cleveref,bm}
\usepackage{graphicx}
\usepackage{siunitx}
\usepackage{pgfplots}
\pgfplotsset{compat=1.14}
\usepgfplotslibrary{statistics}
\usepgfplotslibrary{groupplots}
\usepgfplotslibrary{fillbetween}
\usepackage{tikz}
\usetikzlibrary{positioning}
\usetikzlibrary{decorations.markings}
\usetikzlibrary{decorations.pathmorphing}
\usepackage{url}
\usepackage[utf8]{inputenc}
\usepackage{enumitem}

\theoremstyle{plain}% Theorem-like structures provided by amsthm.sty
\newtheorem{theorem}{Theorem}[section]
\newtheorem{lemma}[theorem]{Lemma}

\theoremstyle{definition}
\newtheorem{definition}[theorem]{Definition}
\newtheorem*{example}{Explanatory example}
\newtheorem*{exampleCont}{Explanatory example (continued)}

\theoremstyle{remark}
\newtheorem{remark}{Remark}

\DeclareMathOperator{\start}{start}

\newcommand{\sched}{1 \, \vert \, r_j, \tilde{d}_j, \text{fixed order} \, \vert \, \Sigma \fEnoarg}

\newcommand{\pA}{\alpha}
\newcommand{\pB}{\beta}
\newcommand{\pC}{\rho}

\newcommand{\paramAlpha}{\gamma}
\newcommand{\paramBeta}{\delta}
\newcommand{\R}{\mathbb R}

\newfont{\sBbb}{msbm8 scaled\magstep1}

\newcommand{\rr}{\mathbb{R}}

\newcommand{\cels}[1]{\SI{#1}{\celsius}}
\newcommand{\kWatt}[1]{\SI{#1}{\kilo\watt}}

\newcommand{\secs}[1]{\SI{#1}{\second}}
\newcommand{\percents}[1]{\SI{#1}{\percent}}

\newcommand{\vecP}{\bm{p}} % Processing times
\newcommand{\vecR}{\bm{r}} % Release times
\newcommand{\vecD}{\bm{\tilde{d}}} % Deadlines
\newcommand{\vecS}{\bm{s}} % Processing times

\newcommand{\cNtasks}{n} % Processing time of #1
\newcommand{\cNblocks}{k} % Number of blocks
\newcommand{\cPmin}{p_{\rm \min}} % Minimal processing time
\newcommand{\cPmax}{p_{\rm \max}} % Maximal processing time
\newcommand{\cP}[1]{p_{#1}} % Processing time of #1
\newcommand{\cR}[1]{r_{#1}} % Release time of #1
\newcommand{\cD}[1]{\tilde{d}_{#1}} % Deadline of #1

\newcommand{\idxTask}{i} % Task index
\newcommand{\idxTaskAnother}{i^{\prime}} % Task index
\newcommand{\idxBlock}{j} % Block index
\newcommand{\idxBlockSize}{m} % Block index
\newcommand{\idxInterval}{k} % Block index

\newcommand{\setTask}{T} % Task set
\newcommand{\setRealNonNeg}{\mathbb{R}_{\geq 0}} % 
\newcommand{\setIntNonNeg}{\mathbb{Z}_{\geq 0}} %
\newcommand{\setIntPos}{\mathbb{Z}_{> 0}} % 

\newcommand{\varS}[1]{s_{#1}} % Task start
\newcommand{\vertex}[2]{v_{#2}^{\,#1}} % vert
\newcommand{\vertS}[1]{\vertex{r}{#1}} % vert start
\newcommand{\vertD}[1]{\vertex{\tilde{d}}{#1}} % vert end
\newcommand{\vertDummyS}{v^{\,s}} % dummy start
\newcommand{\vertDummyE}{v^{\,e}} % dummy end

\newcommand{\fUniform}[2]{\mathit{U}\{#1,#2\}} % Uniform int. distribution
\newcommand{\fExponential}[1]{\mathit{Exp}(#1)} % Exponential distribution
\newcommand{\fstart}[1]{\start\mathopen{}\left(#1\right)\mathclose{}} % Energy function
\newcommand{\fCnoarg}{c} % Energy function
\newcommand{\fC}[1]{\fCnoarg(#1)} % Energy function
\newcommand{\fEnoarg}{E} % Energy function
\newcommand{\fE}[1]{E\left(#1\right)} % Energy function
\newcommand{\fEtotalNoarg}{E_{\text{total}}} % Energy function
\newcommand{\fEtotal}[1]{\fEtotalNoarg\left(#1\right)} % Energy
\newcommand{\symbDelta}{t_f} % idle period length

\newcommand{\fIdle}[2]{\symbDelta\mathopen{}\left(#1,#2\right)} % Length of idle period between two supports

\definecolor{clr1}{RGB}{57,106,177}
\definecolor{clr2}{RGB}{218,124,48}
\definecolor{clr3}{RGB}{62,150,81}
\definecolor{clr4}{RGB}{204,37,41}
\begin{document}

\title{A polynomial-time scheduling approach to minimise idle energy consumption: an application to an industrial furnace}

\author{
\name{Ond\v{r}ej~Benedikt\textsuperscript{a,b}\thanks{Corresponding author~O. Benedikt. Email: benedond@fel.cvut.cz\newline \copyright 2020. This manuscript version is made available under the CC-BY-NC-ND 4.0  \href{http://creativecommons.org/licenses/by-nc-nd/4.0/}{license}.}, %
Baran~Aliko\c{c}\textsuperscript{a}, %
P\v{r}emysl~\v{S}\r{u}cha\textsuperscript{a}, %
Sergej~\v{C}elikovsk\'y\textsuperscript{c} and %
Zden\v{e}k~Hanz\'alek\textsuperscript{a}}
\affil{\textsuperscript{a}Czech Institute of Informatics, Robotics and Cybernetics, Czech Technical University in Prague, Czech Republic}
\affil{\textsuperscript{b}Faculty of Electrical Engineering, Czech Technical University in Prague, Czech Republic}
\affil{\textsuperscript{c} Czech Academy of Sciences, Institute of Information Theory and Automation, Czech Republic}
}

\maketitle

\begin{abstract}
This article presents a novel scheduling approach to minimise the energy consumption of a machine during its idle periods.
In the scheduling domain, it is common to model the behaviour of the machine by defining a small set of machine modes, e.g. ``on'', ``off'' and ``stand-by''.
Then the transitions between the modes are represented by a static transition graph. 
In this paper, we argue that this type of model might be too restrictive for some types of machines (e.g. the furnaces). 
For such machines, we propose to employ the complete time-domain dynamics and integrate it into an idle energy function.
This way, the scheduling algorithm can exploit the full knowledge about the machine dynamics with minimised energy consumption encapsulated in this function.
In this paper, we study a scheduling problem, where the tasks characterised by release times and deadlines are scheduled in the given order such that the idle energy consumption of the machine is minimised.
We show that this problem can be solved in polynomial time whenever the idle energy function is concave. 
To highlight the practical applicability, we analyse a heat-intensive system employing a steel-hardening furnace. 
We derive an energy optimal control law, and the corresponding idle energy function, for the bilinear system model approximating the dynamics of the furnace (and possibly other heat-intensive systems). 
Further, we prove that the idle energy function is, indeed, concave in this case. Therefore, the proposed scheduling algorithm can be used. 
Numerical experiments show that by using our approach, combining both the optimal control and optimal scheduling, higher energy savings can be achieved, compared to the state-of-the-art scheduling approaches.
\end{abstract}


\section{Introduction}

The machines in heat-intensive processes (such as furnaces) are highly energy-demanding, and therefore their energy consumption optimisation usually provides a significant reduction in production costs. In this work, we focus on the idle energy consumption optimisation, which has been widely studied in recent years \citep[see, e.g.][]{2005:Gutowski,2007:Mouzon,2014:Shrouf,2015:Gahm,2017:Che-unrelated,2019:abikarram}.
The research presented in this paper is inspired by a heat-intensive production process from \v{S}koda Auto. There, steel hardening is performed in electric vacuum furnaces, which require high power input to reach and maintain the specific operating temperature.
In this production line, all furnaces are heated to the operating temperature at the beginning of the week and turned off at its end. However, this strategy is very wasteful because a considerable amount of energy is consumed for heating even during the periods when no material is being processed.
The problem of energy-wasting during prolonged idle periods is not specific only to this particular plant. Similar observations have already been made in other companies as well \citep{2007:Mouzon}.

A common approach in the area of the idle energy consumption optimisation is to define a set of machine modes, typically ``off'', ``on'', and ``stand-by'' \citep{2007:Mouzon,2014:Shrouf,2017:Che-unrelated,2019:abikarram}. The feasible transitions between the modes are then represented by a static \emph{transition graph} defining the time and energy needed to switch from one mode to another and thus describing the machine dynamics to some extent.
In this paper, we argue that this type of model might be too restrictive for some types of machines (e.g. the furnaces). 
For such machines, we propose to employ the complete time-domain behaviour of the machine, when available, in contrast to the use of the finite number of stand-by modes as in the existing literature. The relation between the length of the idle period and the possible minimal energy consumption is then represented by the \emph{idle energy function}, which is used by the proposed scheduling algorithm. This way, the whole energy minimisation problem is decomposed into two independent optimisation problems: (i) determination of the idle energy function and (ii) optimal scheduling of the tasks.

For the scheduling part, we examine a single machine problem
where tasks are characterised by release times, processing times, and deadlines while the objective is to minimise the idle energy consumption. Besides, we assume a fixed order of tasks. The reason for this assumption is that the single machine problem with release times and deadlines is already $\mathcal{NP}$-hard \citep{1977:Garey}. 
Therefore, it is reasonable to solve the entire production problem by a heuristic. In this case, a decision concerning the order of tasks and their assignment to machines is often determined by a local-search or meta-heuristic. These techniques can employ the scheduling approach proposed in this paper for finding the optimal start times of the tasks given their order. We prove that whenever the idle energy function is concave, the scheduling problem can be solved in polynomial time by reduction to the shortest path problem. The main advantage of this transformation is that the size of the reduced problem is independent of the length of the scheduling horizon.

The determination of the idle energy function is specific to the considered machine. In this paper, we take as an example electric furnaces widely used in industrial production lines such as steel hardening and glass tempering, operating at a specified temperature. 
Using the Pontryagin's minimum principle (PMP) to analyse a realistic bilinear model of the continuous-time furnace dynamics, we prove that the energy-optimal control law during any idle period is to switch from zero input power (cooling) to the maximum applicable input power (maximal heating) at some convenient switching time. This optimal control law is then shown to result in the concavity of the idle energy function, which enables to employ the proposed optimal scheduling algorithm. The theoretical approach and findings are validated through a case study investigating an industrial furnace in a real production line.

\subsection{Related work}

Concerning the research of energy-efficient manufacturing systems, one of the first analyses in this area was performed by \cite{2007:Mouzon}, who observed that a significant amount of energy could be saved by managing the state of the machine. They proposed several dispatching rules for online production, considering \emph{operating} and \emph{idle} states of the machine. Specifically, rules were devised to turn the non-bottleneck machines off when they were idle for a certain amount of time. Experimental results showed that, compared to the worst-case policy (no switching), substantial energy savings could be achieved. 
This research laid the foundations for further works investigating the minimisation of (idle) energy in production.
Often, following the example of Mouzon et al., authors consider only a simple case with two states, the \emph{processing} (operational) state and \emph{off} state. That is also the case in the work of \cite{2017:Che}, who proposed a mixed-integer linear programming (ILP) model and heuristics for bi-objective minimisation of the energy and maximum tardiness. 
Another example can be found in the work of \citet{2018:Zhou}, who proposed a mathematical model and a differential evolution algorithm for a parallel batch processing machine scheduling problem considering minimisation of the makespan and total energy consumption. Two states of the batch processing machine were assumed for the modelling, namely the \emph{processing} and \emph{idle} state.
\cite{2012:Angel} analysed a single machine problem with tasks characterised by release times and agreeable deadlines and showed that the problem of idle energy minimisation can be solved in polynomial time when only on-off switching is considered. 
Machines characterised by three states (\emph{processing}, \emph{idle}, and \emph{shutdown}) were studied by both \cite{2014:Shrouf} and \cite{2018:Aghelinejad}, who addressed energy minimisation under variable energy prices. 
A common aspect of all previously mentioned works is that the dynamics of the machine is simplified to several constants (representing the transition times/costs between pairs of modes) only. Contrary to that, we show that by using a more precise model of the machine dynamics, higher energy savings can be achieved. Our claim is supported by a case study examining a heat-intensive system employing a steel-hardening furnace.

Regarding scheduling for heat-intensive production systems and industrial furnaces, the literature is still very sparse. Some authors have studied re-heating furnaces \citep{2002:Zhang,2014:Tang}, which are used to heat steel slabs to a specified temperature before they enter the next production stage. Typically, the duration which the slabs spend inside the furnace (i.e. the processing time),
and the sequence of the slabs are optimised. 
\cite{2011:Hait} studied the problem where the metal is melted in several induction furnaces. The melting time can be shortened by increasing the input power. In contrast, the processing time, as well as the temperature, are specified in our case to ensure the desired quality of the product. \cite{2018:Liu} addressed a glass production flow-shop problem, modelling multiple stages, and optimising the makespan and total energy consumption. However, only the \emph{processing} and \emph{idle} states were considered to approximate the furnace dynamics in the scheduling model.

In addition to the manufacturing processes mentioned previously, the research on power-saving states has a broad base in the domain of embedded systems, where energy savings are crucial to prolonging the battery life \citep{2003:Irani, 2007:Baptiste, 2013:Gerards}. The considered devices typically have only a small number of power-saving states \citep{2013:Gerards}, which are specified by the manufacturer. Sometimes authors assume only the \emph{processing state} and the \emph{off-state} \citep{2003:Irani,2007:Baptiste}. The studied problems commonly lead to online scheduling algorithms because of their real-time character or uncertainties in the arrival times of the tasks.
In contrast to embedded systems, the dynamics of machines in production lines, e.g. for the heat-intensive systems investigated in our case study, is typically much slower. Thus, by assuming only \emph{on} and \emph{off} states for such machines, the idle periods between two consecutive tasks would need to be very long to make the transitions possible. 
Another difference is the possibility of solving the production problems offline with respect to known, or \textit{a priori} approximated, parameters of the tasks and the identifiable dynamic behaviours. However, despite all differences, some concepts originating from the domain of embedded systems are general and can still be used even for production scheduling. Frequently, the idle energy consumption is captured by an \emph{idle energy function}, $\fEnoarg : \setRealNonNeg \rightarrow \setRealNonNeg$, mapping the length of the idle period to energy consumption \citep{2013:Gerards}. Such a function $\fEnoarg$ is typically assumed to be non-decreasing piecewise-linear concave where each linear segment corresponds to a single power-saving state. 
Adopting this concept, we mainly propose a new polynomial-time scheduling algorithm, also suitable for production line machines whose dynamics can be captured by a concave idle energy function.

\subsection{Contributions and outline}

The main contribution of this paper is twofold. First, we propose a new polynomial scheduling algorithm using the concept of the idle energy function. Second, we show that the idle energy function can be used to better represent the dynamics of the machine compared to the approaches that are just approximating it with few states only. As the experimental results show, we can achieve much better energy savings. Further, we list the particular contributions of our article in the context of the present related works:
\begin{enumerate}
    \item We define the problem of idle energy consumption minimisation for a single machine scheduling with release times, deadlines, and the fixed order of tasks where the consumption of the machine is defined by the idle energy function (\Cref{sec:problem-statement}).
    \item We suggest decomposing the studied problem to (i) the determination of the idle energy function with respect to the machine dynamics, and (ii) the optimal scheduling of tasks.
    \item We show that the scheduling problem can be solved in $\mathcal{O}(\cNtasks^3)$, where $\cNtasks$ is the number of tasks, assuming that the idle energy function is concave (\Cref{sec:complexity-analysis}). To the best of our knowledge, the closest work that can be adapted to our problem is the algorithm proposed for a fixed sequence of tasks in \citep{2019:Aghelinejad}. The complexity of their algorithm is $\mathcal{O}(|H|^2 \cNtasks)$, where $|H|$ is the length of the scheduling horizon. Since for practical applications $|H| \gg \cNtasks$, our approach exhibits a better complexity (\Cref{sec:experiment-sota-approach}).
    \item Utilising a bilinear system approximation of furnace dynamics, we propose an energy-optimal control law for fixed idle period lengths and show that the idle energy function under this control law is concave (\Cref{sec:modeling}).
    \item Combining the scheduling approach and the idle energy function derived for a real industrial furnace at \v{S}koda Auto (in \Cref{sec:case-study}), we verify the proposed approach on a set of instances and show (in \Cref{sec:exp-1}) that the proposed solution provides significantly less energy consumption as compared with the existing modelling approach based on explicit modelling of the machine modes \citep{2007:Mouzon,2014:Shrouf,2017:Che-unrelated,2019:abikarram}.
\end{enumerate}
\begin{sloppypar}

The rest of the article is organised as follows. \Cref{sec:problem-statement} provides the problem description and assumptions. In \Cref{sec:complexity-analysis}, the dominant structures in schedules are identified, and it is shown that the scheduling problem can be solved in polynomial time by finding the shortest path in a directed acyclic energy graph.
\Cref{sec:modeling} addresses the modelling of the furnace; a bilinear model is described, and the energy-optimal control law is derived. The case study in \Cref{sec:case-study} describes a real furnace used in the production; bilinear model parameters are identified, and the idle energy function is derived. The case study is followed by \Cref{sec:experiments}, which shows the results of the numerical scheduling experiments using the identified model of the real furnace in contrast to the state-of-the-art modelling techniques assuming a finite number of machine modes. Finally, \Cref{sec:conclusions} concludes the article.
\end{sloppypar}

\section{Problem statement} \label{sec:problem-statement}

We study a scheduling problem denoted $\sched$, i.e. the minimisation of the idle energy consumption on a single machine where the order of the tasks is fixed.
Formally, let $\setTask = \{1, 2, \dots, \cNtasks \}$ denote the set of tasks sorted according to the given order. Each task $\idxTask \in \setTask$ is characterised by three integers: release time $\cR{\idxTask} \in \setIntNonNeg$, deadline $\cD{\idxTask} \in \setIntPos$, and processing time $\cP{\idxTask} \in \setIntPos$, such that $\cR{\idxTask} + \cP{\idxTask} \leq \cD{\idxTask} \ \forall \idxTask \in \setTask$.

A schedule is defined by vector of start times $\vecS = (\varS{1}, \varS{2}, \dots, \varS{\cNtasks}) \in \setRealNonNeg^{\cNtasks}$.  A \emph{feasible schedule} is such a schedule that satisfies the following constraints.

\begin{enumerate}[label=(C\arabic*)]
    \item Each task $\idxTask$ is processed within its execution time window $[\cR{\idxTask}, \cD{\idxTask}]$.
    \item The processing order of the tasks is given and fixed.
    \item At most, a single task is processed at one time.
    \item The processing is done without preemption.
\end{enumerate}

\noindent For the rest of this work, when we talk about a schedule, we always mean a feasible schedule.

We assume that the machine is turned \textit{on} (e.g. heated to the operating temperature from \textit{off} state in case of a furnace) just before the first task is processed, and it is turned \textit{off} immediately after the last task is processed. When the machine is \textit{off}, the power consumption is zero. Costs for turning the machine on and shutting it off are constant and cannot be optimised. 

When a task is processed, the machine operates in the processing state given by the respective technological process (e.g. the furnace is heated to the operating temperature, which is the same for all tasks).
Therefore, energy consumption cannot be optimised in this case, as well. However, during the idle periods, the machine can change its state to lower the energy consumption (i.e. the temperature of the furnace can be lowered to save energy). At the end of the idle period, the machine needs to be switched back to the processing state before the next task is processed. 

The objective is to find start times $\vecS$, such that the idle energy consumption $\fEtotal{\vecS}$, i.e. the total energy consumption during idle periods, is minimised. 
An \emph{idle period} is defined as the duration between the completion time of a task and start time of the following one.
Since the execution order of the tasks is fixed, we can assume that the tasks are sorted in the given order, i.e. $\varS{\idxTask} + \cP{\idxTask} \leq \varS{\idxTask+1} \, \forall \idxTask \in \{1,2,\dots, \cNtasks-1\}$.
Then, the objective can be written as
\begin{equation} \label{eq:objective}
    \min_{\vecS} \fEtotal{\vecS} =  \min_{\vecS} \sum\limits_{\idxTask=1}^{\cNtasks-1} \fE{\varS{\idxTask+1} - (\varS{\idxTask} + \cP{\idxTask})},
\end{equation}
where $\fEnoarg : \setRealNonNeg \rightarrow \setRealNonNeg$ represents the \emph{idle energy function}, which encodes the 
relationship between the idle period length and the consumed energy (taking into account various power-savings).
The idle energy function is further discussed in \Cref{sec:control-and-idle-energy}, and a real example for an industrial furnace is shown in \Cref{fig:enery-graphs} in \Cref{sec:case-study}.

Note that because of the fixed order, release times and deadlines can be propagated. Specifically, taking tasks from left to right, release times can be shifted such that 
\begin{equation}
    \cR{\idxTask} := \max \{\cR{\idxTask-1} + \cP{\idxTask-1}, \cR{\idxTask}\}, \ \forall \idxTask \in \{2,3,\dots,\cNtasks\},
\end{equation}
and taking the tasks from right to left, deadlines can be adjusted such that
\begin{equation} \label{eq:propagate-deadlines}
    \cD{\idxTask} := \min \{\cD{\idxTask+1} - \cP{\idxTask+1}, \cD{\idxTask}\}, \ \forall \idxTask \in \{\cNtasks-1, \cNtasks-2, \dots, 1\}.
\end{equation}
If there exists a task such that its propagated execution window is shorter than its processing time, then the instance does not have a feasible solution for the given order. For the rest of this article, we assume that release times and deadlines are propagated and a feasible solution exists.

\section{Scheduling algorithm and complexity analysis} \label{sec:complexity-analysis}

In this section, we show that $\sched$ can be solved in polynomial time under the assumption that the energy function $\fEnoarg$ is concave. Note that if the order was not fixed, the problem would be $\mathcal{NP}$-hard because its underlying problem $1\vert \cR{j}, \cD{j} \vert -$ is $\mathcal{NP}$-complete in a strong sense \citep{1977:Garey}.

A special version of the problem studied here was addressed by \cite{2013:Gerards}, who assumed a so-called \textit{frame-based system}, i.e. a system where $\cR{\idxTask} = (\idxTask-1)\cdot T$ and $\cD{\idxTask} = \idxTask \cdot T$ for some constant number $T$. In frame-based systems, execution windows of the tasks do not overlap. Gerards and Kuper showed that idle energy minimisation in frame-based systems can be done in polynomial time, assuming that the idle energy function is concave. We extend their result to $\sched$, i.e. to systems with arbitrary release times and deadlines, assuming that the execution order of the tasks is fixed. 

Further, we describe the structure of the \emph{energy graph}, and show that $\sched$ can be solved by finding the shortest path in that graph. But first, we provide necessary definitions and show that only schedules in a special form (so-called \emph{block-form schedules}) can be assumed for the optimisation. 

\subsection{Definitions}

A basic structure that appears in the feasible schedules is called a \emph{block of tasks} or simply \emph{block}, and is widely used; see, e.g. \cite{Baker2009} or \cite{2007:Baptiste}.

\begin{sloppypar}
\begin{definition}[Block of tasks]
A sequence of tasks ${B = (b_1,\dots, b_\idxBlockSize)}$, which are scheduled on the same machine, is called a block of tasks if the following properties hold:
\begin{align}
    \varS{b_{\idxTask}} + \cP{b_{\idxTask}} = \varS{b_{\idxTask+1}}, \ \forall \idxTask \in \{1,2,\dots,\idxBlockSize-1\},  \label{eq:block-1} \\
    \forall \idxTask \in \setTask \setminus B: (\varS{\idxTask} + \cP{\idxTask} < \varS{b_{1}}) \vee (\varS{\idxTask} > \varS{b_{\idxBlockSize}} + \cP{b_{\idxBlockSize}}). \label{eq:block-2}
\end{align}
\end{definition}
\end{sloppypar}

Property \eqref{eq:block-2} states that block $B$ is maximal, i.e. it cannot be extended to the left or right. Every feasible schedule is composed of blocks of tasks, which are separated by idle intervals. Blocks are, therefore, fundamental building elements out of which the resulting schedule is created. 

Even though all schedules are composed of blocks of tasks, some schedules are special in a certain sense. We call them \emph{block-form schedules}.

\begin{definition}[Block-form schedule] \label{def:block-form}
A schedule consisting of $\cNblocks$ blocks $B_1$, $B_2$, \dots, $B_{\cNblocks}$ is in the \emph{block form} if each block of tasks $B_\idxBlock$ contains at least one task, which starts at its (propagated) release time or ends at its (propagated) deadline; such a task is called the \emph{support} of block $B_{\idxBlock}$.
\end{definition}

Thanks to the properties of the block-form schedules, the idle energy optimisation can be made simple, as shown in \Cref{sec:dominance-block-form} and \Cref{sec:energy-graph}.

\subsection{Dominance of block-form schedules} \label{sec:dominance-block-form}

In this section, we show that block-form schedules weakly dominate all other schedules. To prove this, we utilise the following lemma.

\begin{lemma} \label{lemma:prolonging-is-good}
Given a concave idle energy function $\fEnoarg: \setRealNonNeg \rightarrow \setRealNonNeg$, for $0 \leq \epsilon \leq x \leq y$ it holds that
\begin{equation} \label{eq:concavity-prolonging}
\fE{x-\epsilon} + \fE{y+\epsilon} \leq \fE{x} + \fE{y}.    
\end{equation}
\end{lemma}

\begin{proof}
Property \eqref{eq:concavity-prolonging} is directly implied by the concavity of $\fEnoarg$, see \cite{2013:Gerards}.
\end{proof}

Lemma \ref{lemma:prolonging-is-good} implies that, in the case of having two idle periods $x$ and $y$, energy $\fE{x} + \fE{y}$ decreases or remains the same even if the shorter idle period of length $x$ is reduced on behalf of the longer idle period of length $y$. Then, we have the following theorem.

\begin{theorem} \label{theorem:block-transformation}
Given a concave idle energy function $\fEnoarg$, for every feasible schedule $S_1$ defined by start times $\vecS_1$, there exists a feasible schedule $S_2$ defined by start times $\vecS_2$, such that $S_2$ is in a block form and $\fEtotal{\vecS_1} \geq \fEtotal{\vecS_2}$.
\end{theorem}

\begin{proof}
If $S_1$ is already in a block form, nothing has to be done. Otherwise, $S_1$ consists of $k$ blocks 
$$ \{B_1, B_2, \dots, B_{\cNblocks} \} = \mathcal{B}_{\text{fixed}} \cup \mathcal{B}_{\text{free}}, \; \mathcal{B}_{\text{fixed}} \cap \mathcal{B}_{\text{free}} = \emptyset,$$
where $\mathcal{B}_{\text{fixed}}$ is the set of blocks that contain at least one support, and $\mathcal{B}_{\text{free}}$ are the blocks without supports. The blocks in $\mathcal{B}_{\text{fixed}}$ will not be moved, while the blocks in $\mathcal{B}_{\text{free}}$ will be shifted to gain a support. By \emph{shift}, we mean adding a non-zero constant to all start times of the tasks in the block.

Let us assume that there is an infinitely long idle period before the first block in $S_1$ and after the last one. Now, every block is separated from the other blocks by two idle periods (before and after the block). 

Let us take an arbitrary block $B \in \mathcal{B}_{\text{free}}$. Since it does not contain a support, it can be shifted. The direction of the shift can be selected according to Lemma \ref{lemma:prolonging-is-good} such that the idle energy consumption does not increase (i.e. shift the block such that the shorter neighbouring idle period decreases its length). Note that the leftmost  (rightmost) block is always shifted right (left) to prolong the time when the machine is off (idle energy consumption does not increase).

After the block is shifted as much as possible, there are two possible outcomes.

\begin{enumerate}
    \item Some task $\idxTask \in B$ reaches its release time or deadline.
    
    In this case, block $B$ gains a support and joins $\mathcal{B}_{\text{fixed}}$; the cardinality of $\mathcal{B}_{\text{free}}$ decreases by one.
    
    \item Block $B$ reaches its neighbouring block $B_{\text{neigh}}$.
    
    In this case, block $B$ joins its neighbouring block. If $B_{\text{neigh}} \in \mathcal{B}_{\text{fixed}}$, then $B$ gains a support and joins $\mathcal{B}_{\text{fixed}}$. Otherwise, $\mathcal{B}_{\text{free}} := (\mathcal{B}_{\text{free}} \setminus \{B, B_{\text{neigh}}\}) \cup \{B \oplus B_{\text{neigh}}\}$, i.e. $B$ and $B_{\text{neigh}}$ are joined (operator~$\oplus$). Anyway, the cardinality of $\mathcal{B}_{\text{free}}$ decreases by one. 
\end{enumerate}

If cases 1. and 2. happen at the same time, both $B$ and $B_{\text{neigh}}$ gain a support, join $\mathcal{B}_{\text{fixed}}$, and the cardinality of $\mathcal{B}_{\text{free}}$ decreases by at least one.

It can be seen that after one shift, the cardinality of $\mathcal{B}_{\text{free}}$ decreases, and the idle energy consumption does not increase (by Lemma \ref{lemma:prolonging-is-good}). By iteratively shifting the blocks without supports, every block will eventually join $\mathcal{B}_{\text{fixed}}$.
Since there are at most $\cNtasks$ blocks in $\mathcal{B}_{\text{free}}$ at the beginning, and the cardinality of $\mathcal{B}_{\text{free}}$ decreases after each shift, $\mathcal{B}_{\text{free}}$ will be empty after at most $\cNtasks$ iterations. Also, there are at most $\cNtasks$ tasks in each block. Therefore, each shift can be done in $\mathcal{O}(\cNtasks)$ steps (shifting one task after another). Hence, the transformation can be done in $\mathcal{O}(\cNtasks^2)$ steps. Schedule $S_2$ is then given by the start times of the tasks in $\mathcal{B}_{\text{fixed}}$.
\end{proof}

\Cref{theorem:block-transformation} shows that it is sufficient to optimise only over schedules in the block form.

\subsection{Finding an energy-optimal block-form schedule} \label{sec:energy-graph}

Here we show how the schedules can be represented as paths in an oriented directed acyclic \emph{energy graph}. The graph-based approach was originally introduced for frame-based systems by \cite{2013:Gerards}, but since the release times and deadlines in their frame-based systems do not overlap, the graph had a very simple structure. In our case, we need to non-trivially extend the idea, relying on \Cref{theorem:block-transformation}.

By \Cref{def:block-form}, each block of a block-form schedule contains at least one support. The main idea leading to a graph-based approach is to represent the supports of the schedule by nodes of the energy graph. In the following, we will show that paths in the energy graph can be associated with the block-form schedules and that the shortest path corresponds to the optimal block-form schedule.

\begin{figure}
    \centering
    \begin{tabular}{ccc}
        \includegraphics[]{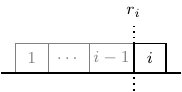} &
        \includegraphics[]{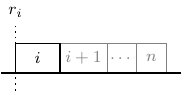} &
        \includegraphics[]{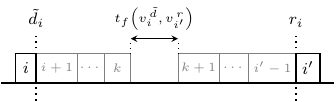} \\
        (a) & (b) & (c) 
    \end{tabular}
    \caption{Examples of the partial schedules corresponding to the edges between (a) $\left(\vertDummyS, \vertex{r}{\idxTask}\right)$, (b) $\left(\vertex{r}{\idxTask}, \vertDummyE\right)$, and (c) $\left(\vertex{\tilde{d}}{\idxTask}, \vertex{r}{\idxTaskAnother}\right)$.}
    \label{fig:edges-examples}
\end{figure}

Our extended version of the energy graph can be represented as a triplet $G = (V_G,E_G,c)$, where $V_G$ is set of its vertices, $E_G$ is set of its oriented edges, and $\fCnoarg: E_G \rightarrow \setRealNonNeg$ is the cost function. For each task $\idxTask \in \setTask$, we define vertices $\vertS{\idxTask}$ and $\vertD{\idxTask}$ representing situations when task $\idxTask$ starts at its release time and ends at its deadline, respectively. Let $\fstart{\vertex{x}{\idxTask}}$ be the actual start time of the task $\idxTask$ represented by vertex $\vertex{x}{\idxTask}$, i.e.
\begin{equation}
    \fstart{\vertex{x}{\idxTask}} = \begin{cases}
        \cR{\idxTask}, & \ \text{ if $x$ is $r$}, \\
        \cD{\idxTask} - \cP{\idxTask}, & \ \text{ if $x$ is $\tilde{d}$}. \\
    \end{cases}
\end{equation}
Furthermore, let us define two additional dummy vertices, the starting vertex $\vertDummyS$ and the ending vertex $\vertDummyE$. We will define the edges in such a way that the paths between $\vertDummyS$ and $\vertDummyE$ represent block-form schedules.
The set of edges $E_G$ consists of three types of edges, ${E_G = E_G^{(1)} \cup E_G^{(2)} \cup E_G^{(3)}}$, where
\begingroup
\allowdisplaybreaks
\begin{align}
    \begin{split}
    E_G^{(1)}  = {} & \Big\lbrace \left(\vertDummyS, \vertex{x}{\idxTask}\right) \ \big| \ \idxTask \in \setTask, x \in \{r,\tilde{d}\} \ \text{such that the partial schedule given by} \\
    & \varS{\idxTask} := \fstart{\vertex{x}{\idxTask}}, \  \varS{\idxTaskAnother} := \varS{\idxTask} - \sum\limits_{k=\idxTaskAnother}^{\idxTask-1} \cP{k} \ \forall \idxTaskAnother \in \{1, 2, \dots, \idxTask-1 \} \ \text{is feasible} \Big\rbrace,  \label{eq:edges-1}
    \end{split} \\
    \begin{split}
    E_G^{(2)} = {} & \Big\lbrace \left(\vertex{x}{\idxTask}, \vertDummyE\right) \ \big| \ \idxTask \in \setTask, x \in \{r,\tilde{d}\} \ \text{such that the partial schedule given by} \\
    & \varS{\idxTask} := \fstart{\vertex{x}{\idxTask}}, \ \varS{\idxTaskAnother} := \varS{\idxTask} + \sum\limits_{k=\idxTask}^{\idxTaskAnother-1} \cP{k} \ \forall \idxTaskAnother \in \{ \idxTask+1, \idxTask+2,\dots, \cNtasks\} \ \text{is feasible} \Big\rbrace, \label{eq:edges-2}
    \end{split}  \\
    \begin{split}
    E_G^{(3)} = {} & \Big\lbrace \left(\vertex{x}{\idxTask}, \vertex{y}{\idxTaskAnother} \right) \ \big| \ \idxTask \in \setTask, \idxTaskAnother \in \setTask, \idxTask < \idxTaskAnother, \ x,y \in \{r, \tilde{d}\} \  \text{and} \\ 
    & \exists k \in \{\idxTask,\idxTask+1,\dots,\idxTaskAnother-1\} \  \text{such that the partial schedule given by} \\
    & \varS{\idxTask} := \fstart{\vertex{x}{\idxTask}}, \ \varS{\idxTaskAnother} := \fstart{\vertex{y}{\idxTaskAnother}}, \\
    & \varS{a} := \varS{i} + \sum\limits_{l=i}^{a-1} \cP{l} \ \forall a \in \{\idxTask+1,\idxTask+2,\dots,k\}, \\
    & \varS{b} := \varS{\idxTaskAnother} - \sum\limits_{l=b}^{\idxTaskAnother-1} \cP{l} \ \forall b \in \{k+1, \dots, \idxTaskAnother-1 \} \ \text{is feasible} \Big\rbrace. \label{eq:edges-3} 
    \end{split} 
\end{align}
\endgroup
In $E_G^{(1)}$, edges connect the starting vertex $\vertDummyS$ and vertex $\vertex{x}{\idxTask}, \; x \in \{r, \tilde{d}\}, \; \idxTask \in \setTask$, associated with task $\idxTask$. Each edge represents the situation when task $\idxTask$ is the support and tasks $\{1, 2, \dots, \idxTask-1\}$ are aligned to the right, joining the block supported by task $\idxTask$, see \Cref{fig:edges-examples}(a).
Similarly, edges in $E_G^{(2)}$ link $\vertex{x}{\idxTask}, \; x \in \{r, \tilde{d}\}, \; \idxTask \in \setTask$, with the ending vertex $\vertDummyE$. Each edge represents the situations when task $\idxTask$ is the support, and tasks $\{\idxTask+1, \idxTask+2, \dots, \cNtasks\}$ are aligned to the left, joining the block supported by $\idxTask$, see \Cref{fig:edges-examples}(b).
Finally, set $E_G^{(3)}$ represents situations when there are two blocks of tasks supported by $\idxTask$ and $\idxTaskAnother$, respectively. All the tasks $\{\idxTask+1,\idxTask+2,\dots,k\}$ are aligned to the left and join the block supported by $\idxTask$ and tasks $\{k+1,k+2,\dots,\idxTaskAnother-1\}$ are aligned to the right and join the block supported by task $\idxTaskAnother$, see \Cref{fig:edges-examples}(c).

Now, we define the cost function $\fCnoarg$. 
We set the costs of edges in $E_G^{(1)}$ and $E_G^{(2)}$ to zero because the tasks represented by these edges are processed without any idle periods.
The costs of edges in $E_G^{(3)}$ correspond to the idle energy consumption between two blocks of tasks. Even though there might be multiple possible ways to schedule the tasks between the two supports, the processing time of each task is assumed to be constant and so the length of the idle period is invariant for a fixed pair of supports. Let us denote the length of the idle period between blocks supported by $\vertex{x}{\idxTask}$ and $\vertex{y}{\idxTaskAnother}$, where $\idxTaskAnother > \idxTask$, by $\fIdle{\vertex{x}{\idxTask}}{\vertex{y}{\idxTaskAnother}}$, defined by
\begin{equation} \label{eq:delta-definition}
    \fIdle{\vertex{x}{\idxTask}}{\vertex{y}{\idxTaskAnother}} = 
        \fstart{\vertex{y}{\idxTaskAnother}} - \left(\fstart{\vertex{x}{\idxTask}} + \cP{\idxTask} \right) - \sum\limits_{k=\idxTask+1}^{\idxTaskAnother-1} \cP{k}.
\end{equation}
Now, the cost function can be defined in the following way:
\begin{equation} \label{eq:edge-cost}
    \fC{e} = 
        \begin{cases}
            0, \ \text{if} \ e \in E_G^{(1)} \cup E_G^{(2)}, \\
            \fE{\fIdle{\vertex{x}{\idxTask}}{\vertex{y}{\idxTaskAnother}}}, \ \text{if} \ e = \left(\vertex{x}{\idxTask}, \vertex{y}{\idxTaskAnother} \right) \in E_G^{(3)}.
        \end{cases}
\end{equation}

\begin{example}
To illustrate the energy graph, let us consider an arbitrary concave idle energy function $\fEnoarg$ and four tasks characterised by parameters given in \Cref{tab:tasks-parameters}. The corresponding energy graph is shown in \Cref{fig:energy-graph}. Each edge $e$ is labelled by its cost $\fC{e}$, defined by \eqref{eq:edge-cost}. 

Note that there is no edge between $\vertDummyS$ and $\vertex{r}{3}$ because if task $3$ started at its release time, it would not be possible to execute the previous tasks without introducing an idle period ($\cD{2} = 40 < 45 = \cR{3}$). But in that case, the previous tasks would form a different block, having its own support. Therefore, edge $(\vertDummyS,\vertex{r}{3})$ does not bring any additional useful information. The situation is similar for other `missing' edges.

\begingroup
\setlength{\tabcolsep}{10pt}
% Tasks parameters
\begin{table}[h]
\tbl{Example task parameters.}
{
    \begin{tabular}{rrrrr} \toprule
    $\idxTask$ & 1 & 2 & 3 & 4\\
    \cmidrule{1-5}
    $\cR{\idxTask}$ &  0 &  15 & 45 & 80 \\ 
    $\cD{\idxTask}$ & 20 &  40 & 70 & 100 \\
    $\cP{\idxTask}$ & 10 &  15 &  5 & 10 \\ \bottomrule
    \end{tabular}
}
\label{tab:tasks-parameters}
\end{table}
\endgroup
\begin{figure}[h]
    \centering
    \resizebox {10cm} {!} {
        \includegraphics{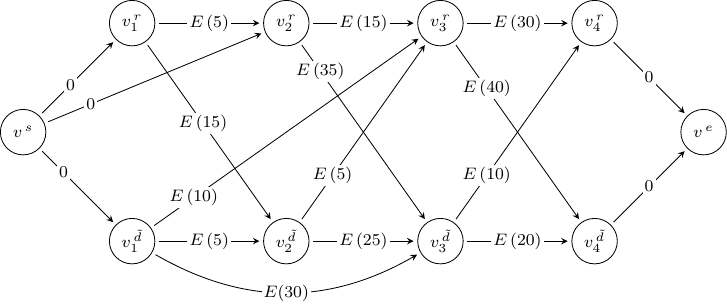}
    }
    \caption{Energy graph constructed for the tasks specified by \Cref{tab:tasks-parameters}.}
    \label{fig:energy-graph}
\end{figure}
\end{example}

The connection between the paths in the energy graph and block-form schedules is explained by the following two lemmas.

\begin{lemma} \label{lemma:paths-1}
For every block-form schedule $S$, there exists a path in the corresponding energy graph, such that length of the path equals the idle energy consumption of schedule $S$.
\end{lemma}

\begin{proof}
This is assured by the structure of the energy graph. Given a block-form schedule with blocks $B_1, B_2, \dots, B_{\cNblocks}$ and their supports $a_1, a_2, \dots, a_{\cNblocks}$, the corresponding path in the energy graph is \hbox{$\vertDummyS, \vertex{x(a_1)}{a_1}, \vertex{x(a_2)}{a_2}, \dots, \vertex{x(a_k)}{a_k}, \vertDummyE$}, where 
\begin{equation}
  x(a_i) := 
    \begin{cases}
        r & \ \text{if } \ a_i \ \text{starts at its release time}, \\
        \tilde{d} & \ \text{if} \ a_i \ \text{ends at its deadline}.
    \end{cases}
\end{equation}
Nodes on the path correspond to the supports of the individual blocks, and because the cost of each edge directly corresponds to the idle energy consumption, the length of the path is the same as the idle energy consumption of the schedule.
\end{proof}
\begin{lemma} \label{lemma:paths-2}
For every path $P$ between the start node $\vertDummyS$ and end node $\vertDummyE$ in the energy graph, there exists a feasible block-form schedule $S$, such that the idle energy consumption cost of $S$ is the same as the length of path $P$. 
\end{lemma}
\begin{proof}
Again, this is trivially given by the structure of the energy graph, where nodes represent supports of the blocks. According to \eqref{eq:edges-1}--\eqref{eq:edges-3}, an edge between two nodes representing the supports is added only if there exists a feasible schedule of the tasks between them.
\end{proof}

Finally, by Lemmas \ref{lemma:paths-1} and \ref{lemma:paths-2}, we see that problem $\sched$ can be solved by finding the shortest path in a directed acyclic graph. The graph contains $\mathcal{O}(\cNtasks)$ vertices and at most $\mathcal{O}(n^2)$ edges. Whether edge $e$ belongs to the graph or not can be verified according to \eqref{eq:edges-1}--\eqref{eq:edges-3} in linear time $\mathcal{O}(\cNtasks)$. Therefore, the number of steps needed to build the graph is upper bounded by $\mathcal{O}(\cNtasks^3)$. 
The shortest path itself can be found in linear time with respect to the size of the graph by the dynamic programming \citep[sec. 24.2]{2001:Cormen}.
So the overall complexity is bounded by $\mathcal{O}(\cNtasks^3)$.

\begin{exampleCont}
The schedule corresponding to path $\vertDummyS, \vertex{\tilde{d}}{1}, \vertex{r}{3}, \vertex{r}{4}, \vertDummyE$ is depicted in \Cref{fig:schedule}. It consists of three blocks, $B_1 = (1,2)$, $B_2 = (3)$, and $B_3 = (4)$. Supports of these blocks are tasks $1$, $3$ and $4$, respectively. Idle energy consumption of the schedule equals the sum of energy consumed during the first idle period (from time 35 to time 45), plus energy consumed during the second idle period (from time 50 to time 80).
\begin{figure}[h]
    \centering
    \resizebox {8cm} {!} {
        \includegraphics{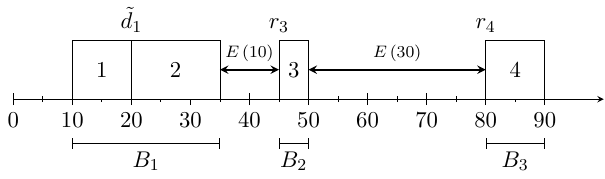}
    }
    \caption{Feasible schedule corresponding to path $\vertDummyS, \vertex{\tilde{d}}{1}, \vertex{r}{3}, \vertex{r}{4}, \vertDummyE$.}
    \label{fig:schedule}
\end{figure}

\end{exampleCont}

\begin{remark}
Note that edges in $E_G^{(3)}$ might not imply one particular schedule of the tasks between the supports. Therefore, for a given path, there might exist multiple feasible schedules with the same idle energy consumption. Similarly, as each block might contain multiple supports, there might be multiple different paths corresponding to one block-form schedule.
\end{remark}

\begin{remark}
The graph-based approach described above can handle arbitrary concave idle energy function, which is a common shape of the idle energy function used in the literature \citep{2003:Irani,2013:Gerards}. However, it is still an open question if the problem would be polynomial even if the idle energy function was not concave but arbitrary.
\end{remark}

\begin{remark}
The energy graph could also be used to find the schedules minimising the number of idle periods longer than 0. Such an application is useful when the stress of the machine caused by excessive switching needs to be minimised. The problem reduces again to the shortest path problem. The structure of the graph remains the same, but the edges in $E_G^{(3)}$ should be labelled by some positive constant, e.g. 1. Note that it is again possible to optimise only over the block-form schedules because the shifts described in the proof of \Cref{theorem:block-transformation} might join some blocks but never split them.
\end{remark}

\section{Electric furnaces: modelling, optimal control and energy function} \label{sec:modeling}

Up till now, we have discussed how to solve scheduling problem $\sched$, assuming that the energy function is given and concave.
The majority of the existing papers addressing the idle energy optimisation assume that the dynamics of the machine is described by a static transition graph, and its parameters are given. Obtaining those parameters or the idle energy function can be simple in some cases (e.g. for some hardware components in the embedded systems, the parameters or the idle energy function can be extracted from the data provided by the manufacturer), but becomes quite challenging in others. Since the idle energy optimisation aims at a large variety of machines ranging from processors to huge furnaces, it is not possible to provide a single approach for obtaining the parameters of the transition graph or the idle energy function. Therefore, we concentrate on heat-intensive systems that are the most frequently addressed in connection with the idle energy optimisation in production.

In this section, we discuss the electric furnace models and present a bilinear modelling approach, which is shown to provide a good approximation of industrial electric furnace dynamics. Further, the open-loop control for minimum energy consumption during idle periods, concerning the studied scheduling problem, is given based on the considered bilinear system approximation. Then, we show that the idle energy function as an input to the scheduling problem is concave under the proposed approximation and control, thus confirming the use of the above-proposed algorithm is correct.

\subsection{A bilinear model approximation of furnaces} \label{sec:furnace-model}

Obtaining and identifying a reasonable physical model of an industrial furnace is usually very difficult due to unspecified characteristics, imperfections or degradation of insulation materials, and time/temperature dependency of the physical parameters.
Thus, instead of proposing a physical model and identifying its parameters, it is usual in practice to approximate the furnace dynamics with reasonable linear and nonlinear mathematical models; see, e.g. \cite{Wang2002} for a linear model, \cite{Moon2003} for a fuzzy system approximation, \cite{Wang1998} for a direction-dependent model, and \cite{2000:Yu, 2008:Chook} for bilinear system approximations.

\begin{sloppypar}
Our decision to use the bilinear approximation of the furnace dynamics is motivated by the existing literature. For example, \cite{Derese1980} have reported that the bilinear model for heat-transfer processes is more suitable than the linear model. \cite{2008:Chook} considered the identification of a first-order bilinear model for an electric tube furnace and showed experimentally that the bilinear model provides the most accurate description as compared with the linear and direction-dependent models. Another advantage of the bilinear model is its simplicity and well-understood behaviour in the class of nonlinear systems. Thus, we also consider the approximation of the furnace dynamics similarly as in \cite{2008:Chook} with the bilinear model
\begin{equation} \label{eq:sys}
    \dot{x}(t) = - \pA x(t) + \pB u(t) - \pC x(t) u(t), \quad x(t)\in \rr, \quad u(t)\in[0,\bar u], \quad \pA,\,\pB,\,\pC\,\in \rr_{>0}
\end{equation}
where $u$ is the applied electric power (in kilowatts), i.e. the input to the system, and $x$ is the deviation of the furnace temperature $T_f$ (in kelvins) from the constant ambient temperature $T_e$, $x(t) := T_f(t) - T_e $, i.e.  the variable to be controlled. The model (\ref{eq:sys}) slightly differs from that in \cite{2008:Chook}, because we additionally accommodate constraints on control and system parameters regarding the reality for furnaces. First, we do impose the upper bound $\bar u$ on the admissible control power, which is important in practice. Second, based on physical modelling considerations, it is assumed in \eqref{eq:sys} that the system parameters $\pA$, $\pB$, and $\pC$ are positive constants. That is due to \cite{2008:Chook}, where Section IV provides successful identification of $\pA, \pB, \pC$, resulting as positive numbers for their furnaces operational data. Note, that physical-principle-based modelling provided in \cite{2008:Chook} actually gives the following model
\begin{equation} \label{eq:physical-sys}
    \dot{T}_f(t) = \frac{1}{C_f} \left( - \frac{T_f(t)-T_e}{R} + u(t) - K (T_f^4(t)-T_e^4) \right),
\end{equation}
where $C_f$ is the thermal capacitance, $R$ is the thermal resistance, and $K$ is a constant regarding the emissivity of the furnace.
Obviously, $C_{f},\;R,\;K > 0$; as already noted, $T_{e}$ stands for the ambient temperature, which is assumed constant since its possible variations are negligible compared to extremely high furnace temperatures.
Due to its complexity, instead of \eqref{eq:physical-sys} \cite{2008:Chook} study simpler bilinear model \eqref{eq:sys} and provide some arguments for such a simplification. Indeed, there is a kind of trade-off: higher-order nonlinearity of \eqref{eq:physical-sys} is replaced by bilinear dependence in \eqref{eq:sys}, so rigorously \eqref{eq:sys} is not a simplification or approximation of \eqref{eq:physical-sys}. Yet, as shown in the sequel, \eqref{eq:sys} can be handled in an easier way, and some rigorous mathematical statements can be proved for it. Besides easier theoretical analysis, another argument justifying replacement of \eqref{eq:physical-sys} by \eqref{eq:sys} given in \cite{2008:Chook} is that the nonlinearities that arise in heat-transfer processes may be represented by characteristics that are similar to those of a bilinear system. In such a way, the current paper joins the existing literature mainstream represented by \cite{2008:Chook} and will concentrate on the model \eqref{eq:sys} only. Note, that the constraints on $\pA, \; \pB, \; \pC$ will hold for the electrical vacuum furnace, which is studied in \Cref{sec:case-study} as a case study. As shown in \Cref{sec:paramestimate} later on, these parameters $\alpha, \; \beta, \; \rho $ can be quite precisely identified based on the real data, and the resulting estimates comply with the above assumptions.
\end{sloppypar}

\subsection{Solving the ordinary differential equation with a discontinuous right-hand side}

Before formulating the main theorem of this section analysing the optimal control of system  \eqref{eq:sys}, let us briefly recall the definition of the solution of the ordinary differential equation (ODE) with the possibly discontinuous right-hand side. This  overview is presented in a rather casual way; rigorous and detailed theory can be found, e.g. in \cite[Chapter~1]{1988:Filippov}. Indeed, as it will be seen, the optimal control is a discontinuous function in time and thereby after substituting it into \eqref{eq:sys} one gets ODE with discontinuous (in time variable) right-hand side. Namely, consider ODE
\begin{equation} \label{eq:ode-general}
{\dot x}(t)=f(x(t),t), ~~x\in \rr^{n}.
\end{equation}
The usual definition of the solution of \eqref{eq:ode-general} for its \emph{continuous} right-hand side $f(x,t)$ is that the solution $x(t)$ is a continuously differentiable function of time converting the above ODE into equality valid for all times. As there are infinitely many such solutions, the specific unique solution is determined by the so-called \emph{initial condition}
\begin{equation} \label{eq:ode-initial-condition}
x(t_{0})=x_{0},~~x_{0}\in \rr^{n}, ~t_{0}\in \rr,    
\end{equation}
where $t_{0},x_{0}$ are given initial time and initial condition, respectively. The
relations \eqref{eq:ode-general} and \eqref{eq:ode-initial-condition} are usually referred to as the \emph{initial value problem}, or \emph{Cauchy problem}.
When the right-hand side of \eqref{eq:ode-general} is discontinuous, the solution of \eqref{eq:ode-general} cannot be continuously differentiable in time. When the discontinuity is with respect to time only, the usual way to handle this situation is to define the solution in \emph{Caratheodory sense}; namely, the initial value problem \eqref{eq:ode-general} and \eqref{eq:ode-initial-condition} is replaced by the following integral equation
\begin{equation} \label{eq:ode-integral}
x(t)=x(t_{0}) + \int_{t_{0}}^{t} f(x(\tau ), \tau ) {\rm d}\tau ,
\end{equation}
where the solution $x(t)$ is required to be continuous only. Note, that the solution of the integral equation \eqref{eq:ode-integral} automatically satisfies the initial condition \eqref{eq:ode-initial-condition} and, moreover, where $x(t)$ is in addition continuously differentiable, it implies the validity of \eqref{eq:ode-general}. As already noted, Caratheodory approach helps to handle the discontinuity with respect to the time variable only. The discontinuity with respect to state variable $x$ presents even more tough challenge and even more abstract solution is required, namely the so-called solution in the Fillipov's sense.

In the subsequent analysis, all the time discontinuities will be of the simplest kind, i.e. they will be piecewise continuous. In this case, Caratheodory solution can also be obtained in the following intuitively clear way. Namely, ODE is solved together with the initial condition on the largest time interval where $f(x,t)$ is continuous. When reaching discontinuity point $t_{dc}\in \rr$, the resulting solution value $x(t_{dc})$ is taken as the initial condition for the next time interval where $f(x,t)$ is continuous; ODE is solved again and this procedure can be repeated.

Note that such an approach correctly represents reality. In the case of furnace heating, it means that discontinuous jump change of heating influences further development of the temperature, but the temperature has to stay continuous even at the point where heating intensity experiences jump, see \Cref{fig:optcont}. Obviously, such an understanding of the solution of the ODE with time discontinuity at its right-hand side is the only acceptable one from the natural and practical point of view. Putting it in different words, under quite mild and reasonable mathematical technical assumptions imposed on the right-hand side $f(x,t)$, there is a unique solution that satisfies ODE in a classical sense everywhere except some isolated time moments, where this unique solution is at least continuous. In other words, many solutions are possible, but only one of them is everywhere at least continuous. 

In the sequel, we will use exactly the latter approach to obtain the unique solution of the initial value problem when heating intensity (the input) is piecewise constant. Namely, we compute the solution to the initial value problem on time subinterval where heating intensity is constant. Then, at the time where heating intensity jumps to a different constant value, we use the terminal value of temperature on the first time subinterval as the initial condition for the ODE solution on the next time subinterval.

\subsection{Minimum-energy control and the related idle energy function} \label{sec:control-and-idle-energy}

This subsection aims to study the optimal control of furnaces during an idle period, based on the approximate bilinear model \eqref{eq:sys}. 

Recall that our aim is to find an energy-efficient behaviour of the furnace in an idle period. Thus, we look for an optimal control law, which minimises the power consumption for any fixed idle period length. Then, our problem for furnaces turns into finding a control minimising the performance index
\begin{equation} \label{eq:pi}
    J(u) = \int_{0}^{t_f}{|u(t)|dt}
\end{equation}
which is called as minimum-control-effort problem \citep{kirk2004}. Obviously, $t_f$ can be considered as the idle period length, i.e. \hbox{$(s_{i+1} - (s_i + p_i))$} in \eqref{eq:objective}. Then $J(u)$ is the energy (in kilowatt-hours) consumed during the corresponding idle period, i.e. \hbox{$E(s_{i+1} - (s_i + p_i))$} in \eqref{eq:objective}. Note, that it is sufficient to consider an open-loop control to heat the furnace
to the (close neighbourhood of) operating temperature at the end of the idle period (assuming constant ambient temperature), whereas a closed-loop control is necessary to maintain the operating temperature.
Such a control strategy is actually common in process control applications, e.g. see \Cref{fig:data-600} with the temperature data of the real industrial furnace controlled to operate at different temperatures in our case study. As we seek a control minimising energy consumption during the idle periods, we give the following theorem for the open-loop optimal control problem for the industrial furnaces which can be modelled as the bilinear system in (\ref{eq:sys}).
\begin{theorem} \label{theorem:optcont} 
Consider the following optimal control problem: minimise the performance index (\ref{eq:pi}) 
subject to constraints
\begin{equation}
x(0)=x(t_f)=x_0 \in \R, ~x_{0}>0,
\label{eq:constraints}
\end{equation}
where $x(t)$ is the solution of the system \eqref{eq:sys} and $t_f>0$ is a given fixed terminal time.  
Further, assume that
\begin{equation}
(\pB-\pC x_0)\bar u -\pA x_{0} > 0,
\label{eq:assmp}
\end{equation}
where $\bar{u}$ is the upper bound on $u(t)$.
Then there exists the unique optimal control $u^{*}(t)$ solving the above-defined optimal control problem and this optimal control takes the following form
\begin{equation}
u^{*}(t) = \begin{cases}
0, \quad \forall t \in [0,t_{sw}) \\
\bar{u}, \quad  \forall t \in [t_{sw}, t_f],
\end{cases}
\label{eq:optcont}
\end{equation}
where $t_{sw} \in (0,t_{f})$ is the switching time. Finally, $t_{sw}$ is the solution of the following equation
\begin{equation}
x_0 = \mathrm{exp} \left((-\pA-\pC \bar{u})(t_f-t_{sw})\right) \left( x_0 \; \mathrm{exp} (-\pA t_{sw}) - \frac{\pB \bar{u}}{\pA + \pC \bar{u}} \right) + \frac{\pB \bar{u}}{\pA+\pC\bar{u}},
\label{eq:tsw}
\end{equation}
this solution exists and is unique for any given $t_{f}>0$.
Furthermore, defined in such a way function $t_{sw}(t_{f})$ satisfies
\begin{equation}
\label{implicite}
\frac{{\rm d} t_{sw}}{{\rm d} t_{f}}=1-\frac{\pA x_{0}}{(\pB\,{\rm exp}(\pA t_{sw}) - \pC x_{0})\bar{u}}.
\end{equation}
\end{theorem}
\begin{proof}
Pontryagin's minimum principle (PMP) is used \citep{kirk2004}. To do so, realise that $|u(t)|$ in \eqref{eq:pi} can be replaced simply by $u(t)$ because $u(t)>0 \, \forall t$ in \eqref{eq:sys}. Further, the appropriate Hamiltonian function for the performance index  (\ref{eq:pi}) and the  system (\ref{eq:sys}) is given by
\begin{equation}
H(x(t),u(t),\psi(t)) = u(t) - \pA \psi(t) x(t) + \psi(t) [\pB-\pC x(t)] u(t)
\label{eq:hamil}
\end{equation}
where $\psi(t)$ represents the usual adjoint variable. By PMP, the necessary conditions for $u^{*}(t)$ to be an optimal control are
\begin{subequations}
\begin{equation}
\dot{x}^{*}(t) = \frac{\partial H(x^{*},u^{*},\psi^{*})}{\partial \psi} =  - \pA x^{*}(t) + \pB u^{*}(t)- \pC x^{*}(t)u^{*}(t),
\label{eq:necx}
\end{equation}
\begin{equation}
\dot{\psi}^{*}(t) = -\frac{\partial H(x^{*},u^{*},\psi^{*})}{\partial x} =  \psi^{*}(t)(\pC u^{*}(t)+\pA),~~\psi(0)=\psi_{0}\in \R\setminus \{ 0 \} ,
\label{eq:necp}
\end{equation}
\begin{equation}
H(x^{*}(t),u^{*}(t),\psi^{*}(t)) = \min _{u\in [0,\bar{u} ]} H(x^{*}(t),u(t),\psi^{*}(t)) ~\forall t\in [0,t_{f}] ~~ \Rightarrow
\label{eq:nech1}
\end{equation}
\begin{equation}
u^{*}(t) + \psi^{*}(t) [\pB-\pC x^{*}(t)] u^{*}(t) =  \min _{u\in [0,\bar{u} ]}\left(  u(t) + \psi^{*}(t) [\pB-\pC x^{*}(t)] u(t) \right)  ~\forall t\in [0,t_{f}]  .
\label{eq:nech2}
\end{equation}
\end{subequations}
Indeed, the boundary conditions (\ref{eq:constraints}) of the investigated control problem are fixed, so that $\psi(t)$ can be any nontrivial solution of the adjoint equation (\ref{eq:necp}).

Before analysing the above necessary condition for the optimality, let us give the following property useful later on. Namely, (\ref{eq:necx}) and (\ref{eq:necp}) can be solved analytically giving that
\begin{equation}
x^{*}(t) = \text{exp}\left( -\pA t -\pC \int_{0}^{t}{u^{*}(\eta)d\eta} \right) \left( x_0 + \pB \int_{0}^{t}{\text{exp} \left( \pA \eta + \pC \int_{0}^{\eta}{u^{*}(s) ds}  \right) u^{*}(\eta)d\eta} \right),
\label{eq:solx}
\end{equation}
\begin{equation}
\psi^{*}(t) = \psi_0 \, \text{exp} \left( \pA t+\pC\int_{0}^{t}{u^{*}(\eta)d\eta} \right) .
\label{eq:solp}
\end{equation}
To analyse \eqref{eq:necx}--\eqref{eq:nech2} subject to the control constraint $u(t)\in[0,\bar{u}]$, consider  the function
\begin{equation}
\phi (\psi^{*}(t),x^{*}(t)) = \psi^{*}(t) (\pB -\pC x^{*}(t)) + 1
\label{eq:cond}
\end{equation}
to investigate the minimum of the Hamiltonian with respect to $u$. Further, realise that the necessary condition \eqref{eq:nech1}--\eqref{eq:nech2} implies  that $u(t)=\bar u$ if $\phi (\psi^{*}(t),x^{*}(t)) < 0$;  $u(t)=0$ if $\phi (\psi^{*}(t),x^{*}(t)) > 0$; whereas for $\phi (\psi^{*}(t),x^{*}(t)) = 0$ it is always satisfied. As a consequence, the optimal control, if it exists, satisfies
\begin{equation}
u^{*}(t)  \begin{cases}
= \bar u, & \text{for} \quad \phi (\psi^{*}(t),x^{*}(t)) < 0 \\
= 0, & \text{for} \quad \phi (\psi^{*}(t),x^{*}(t)) > 0 \\
\in [0, \bar u ], & \text{for} \quad \phi (\psi^{*}(t),x^{*}(t)) = 0.
\end{cases}
\label{eq:formopt}
\end{equation}
Furthermore, by \eqref{eq:solx} and \eqref{eq:solp} it holds that
\begin{equation}
\begin{split}
     \phi (t) = 1 - \psi_0 x_0 \pC + \psi_0 \pB \, \text{exp} \left( \pA t+\pC\int_{0}^{t}{u(\eta)d\eta} \right) \\
     - \psi_0 \pB\pC \int_{0}^{t}{\text{exp} \left( \pA t+\pC\int_{0}^{\eta}{u(s)ds}\right) u(\eta)d\eta },  \\
     \frac{d\phi (t)}{dt}  = \psi_0 \pB \, (\pA+\pC u(t)) \, \text{exp} \left( \pA t+\pC\int_{0}^{t}{u(\eta)d\eta} \right) \\
     - \psi_0  \pB \pC \, u(t) \, \text{exp} \left( \pA t+\pC\int_{0}^{t}{u(\eta)d\eta} \right),
\end{split}
\label{eq:phi}
\end{equation}
which implies
\begin{equation}
\frac{d\phi (t)}{dt}  = \psi_0 \pA \pB \, \text{exp} \left( \pA t+\pC\int_{0}^{t}{u(\eta)d\eta} \right).
\label{eq:derphi}
\end{equation}
Now, using  (\ref{eq:phi}) and (\ref{eq:derphi})  one concludes that
\begin{equation}
\phi(0) = \psi_0 (\pB - \pC x_0) + 1 ,
\label{eq:phi0}
\end{equation} \vspace{-12pt}
\begin{equation}
\mathrm{sign}\left(\frac{d\phi}{dt}\right) = \mathrm{sign} (\psi_0), \; \psi_0\neq 0.
\label{eq:signderphi}
\end{equation}
Note that by \eqref{eq:signderphi} $\phi(t)$ is obviously a strictly monotonous function. In such a way, $\phi (t)$ either vanishes at a single isolated point only, or it never vanishes. As $\psi_0\neq 0 $, only the following four options are possible for $u^{*}(t) $ to be optimal.
\begin{enumerate}
	\item If $\psi_0 > (\pC x_0-\pB)^{-1} > 0$, then $\phi(0)>0$ and $\frac{d\phi (t)}{dt}>0$, $\forall t\geq 0$, which means $\phi(t)>0$, $\forall t\geq 0$. By \eqref{eq:formopt}, then $u^{*}(t)\equiv 0$. However, it is clear from \eqref{eq:solx} that \eqref{eq:sys} with $u(t)\equiv u^{*}(t)\equiv 0$ does not satisfy (\ref{eq:constraints}).
	\item If $(\pC x_0-\pB)^{-1} > \psi_0 > 0$, then $\phi(0)<0$ and $\frac{d\phi (t)}{dt}>0$, $\forall t\geq 0$.  By \eqref{eq:formopt}, then $u^{*}(t) = \bar{u}, \; t < t_{sw}$ and $u^{*}(t) = 0, \; t > t_{sw}$. However, this option is not possible because $(\pC x_0-\pB)>0$ contradicts 
	the assumption \eqref{eq:assmp} as $\pA$, $\bar u $ and $x_0$ are positive.
	\item If $\psi_0 < (\pC x_0-\pB)^{-1} < 0$, then $\phi(0)<0$ and $\frac{d\phi (t)}{dt}<0$, $\forall t\geq 0$, which means $\phi(t)<0$, $\forall t\geq 0$. By (\ref{eq:formopt}), then $u^{*}(t)\equiv \bar{u}$. However, by  assumption (\ref{eq:assmp}) and by (\ref{eq:solx}) it holds that  $x(t_f)>x_0$. Thus,  (\ref{eq:constraints}) is violated.
	\item If $(\pC x_0-\pB)^{-1} < \psi_0 < 0$, then $\phi(0)>0$ and $\frac{d\phi (t)}{dt}<0$, $\forall t\geq 0$. By \eqref{eq:formopt}, then
	\begin{equation}
	u^{*}(t) = 0, \; t < t_{sw}, \; u^{*}(t) = \bar{u}, \; t > t_{sw}; \quad t_{sw} = \pA^{-1} \; \mathrm{log} \left( (\pC x_0 \psi_0-1)/(\pB \psi_0)\right).
	\label{eq:optcont2}
	\end{equation}
Moreover, it can be seen through some straightforward analysis that when $\psi_0$ ranges through $((\pC x_0-\pB)^{-1},0)$, the expression $\left( (\pC x_0 \psi_0-1)/(\pB \psi_0) \right)$ ranges through $(1,\infty)$, i.e. $\psi_0$ can always be chosen in such a way that any $t_{sw}\in(0,\infty)$ is possible.
\end{enumerate}
Summarising, the control satisfying PMP and (\ref{eq:constraints}) under assumption (\ref{eq:assmp}) should have the form (\ref{eq:optcont2}) for some suitable switching time $t_{sw}$. To conclude the proof, it  remains to show that there is a unique $t_{sw}\in [0,t_{f}) $ such that  (\ref{eq:sys}) with $u(t)\equiv u^{*}(t)$ given by (\ref{eq:optcont2}) satisfies the boundary conditions  (\ref{eq:constraints}). Such a property follows straightforwardly by (\ref{eq:solx}) and (\ref{eq:assmp}), moreover,  also by (\ref{eq:solx}), the switching time  $t_{sw}$ is the solution of
\begin{equation}
x_0 = {\mathrm{exp}} \left((-\pA-\pC\bar{u})(t_f-t_{sw})\right) \left( \text{exp} (-\pA t_{sw})x_0 - \frac{\pB \bar{u}}{\pA+\pC\bar{u}} \right) + \frac{\pB \bar{u}}{\pA+\pC\bar{u}}.
\label{eq:tsw1}
\end{equation}

Indeed, on the right-hand side of \eqref{eq:tsw1} there is a value of temperature
trajectory $x(t)$ at time $t_{f}$ obtained by  solving \eqref{eq:sys} on
subinterval $[0, t_{sw})$ with initial condition $x(0)=x_{0}$ applying
the input (applied power) $u \equiv 0$ and then solving \eqref{eq:sys} with
initial condition $x(t_{sw}) = {\rm exp}(-\pA t_{sw})x_{0} $ and the input $ u \equiv
{\bar u}$ on subinterval $[t_{sw},t_{f}]$.

Note, that $t_{sw}$ solving (\ref{eq:tsw1}) exists and is unique for any given $t_{f}>0$. Indeed, the right-hand side of (\ref{eq:tsw1}) is a smooth function of $t_{sw}$ and it is equal to  ${\mathrm{exp}} \left( -\pA t_{f} \right) x_{0}<x_{0}$ if $t_{sw}=t_{f}$ and to
$$
{\mathrm{exp}} \left( (-\pA-\pC\bar{u})t_{f} \right) \left( x_0 - \frac{\pB \bar{u}}{\pA+\pC\bar{u}} \right) + \frac{\pB\bar{u}}{\pA+\pC\bar{u}} > x_{0},
$$
if $t_{sw}=0$. The last inequality straightforwardly holds thanks to the assumption (\ref{eq:assmp}) and ${\mathrm{exp}}\left( (-\pA-\pC\bar{u})t_{f} \right)\in (0,1)$. As a consequence, there exists at least one $t_{sw}$ solving (\ref{eq:tsw1}) thanks to the well-known basic property of continuous functions. To show that such $t_{sw}$ is unique, note that the right-hand side of \eqref{eq:tsw1} is strictly decreasing function of $t_{sw}$ since its derivative with respect to $t_{sw}$ is
$$
\bar{u} \cdot {\rm exp}((-\pA -\pC \bar{u})(t_{f}-t_{sw})) \cdot ( \pC x_{0} {\rm exp}(-\pA t_{sw} )  - \pB ),
$$
which is negative since by the assumption \eqref{eq:assmp} $\pB > \pC x_{0} $ and
obviously \hbox{$\pC x_{0} > \pC x_{0} {\rm exp}(-\pA t_{sw})$} as
$\pA > 0, t_{sw} \ge 0$.
In such a way, the value $t_{sw}$ solving \eqref{eq:tsw1} exists and is unique. Finally, to prove \eqref{implicite} apply the well-known formula to compute the derivative of the implicitly defined
function and perform  some straightforward, though laborious
computations. The proof is complete.
\end{proof}
\begin{remark} \label{rem:assmp}
The assumption (\ref{eq:assmp}) is equivalent to $\pA x_0/(\pB -\pC x_0)\in (0,\bar{u})$. The value $\pA x_0/(\pB-\pC x_0)$ is the constant trim control keeping the state $x_{0}$  as the equilibrium, {i.e.} $x(t)\equiv x_0$ and therefore the assumption (\ref{eq:assmp}) should be valid in any reasonable practical setting. Indeed, if the assumption (\ref{eq:assmp}) is to be replaced by
 $(\pB-\pC x_0)\bar u -\pA x_{0} = 0$, then the optimal control is $u^{*}(t) = \bar u, \; \forall t\in [0,t_f]$, {\it{i.e. as if}} $t_{sw}=0$ {\it{in}} (\ref{eq:optcont2}). As such, ${\bar u}=\pA x_0/(\pB -\pC x_0)$ is the trim control value that ensures $x(t)\equiv x_0$; practically, such a situation is not acceptable because any small perturbation pushing the state to a value slightly lower than $x_{0}$ cannot be compensated for.
\end{remark}
\begin{remark} \label{rem:constraints}
We consider the optimal control law with the state constraint (\ref{eq:constraints}) because a single operating temperature $x_0$ for the scheduling problem is considered. Definitely, the furnace temperature is $x_0$ at the beginning of each idle period and should also be $x_0$ at the end of the idle period to execute the consecutive task. In fact, \Cref{theorem:optcont} can be easily extended to a more general case with  boundary conditions of the form $x(0)=x_{0}, \, x(t_{f})=x_{f}, \, x_{0}>0, \, x_{f}>0$ and, possibly, $x_{0}\neq  x_{f}$.
\end{remark}

\begin{sloppypar}
Let us finally show that the energy function of the idle period length, for a furnace described by the bilinear model (\ref{eq:sys}) and optimally controlled as proposed in \Cref{theorem:optcont}, is concave. 
\end{sloppypar}
\begin{theorem}\label{theorem:concavity}
The idle energy function $E: \mathbb{R}_{\geq0} \rightarrow \mathbb{R}_{\geq 0}$ of system (\ref{eq:sys}) under control (\ref{eq:optcont}) assuming (\ref{eq:assmp}) is described by equation $E(t_f) = \bar{u} \cdot (t_f - t_{sw}(t_f))$ for any $t_f \in \mathbb{R}_{\geq0}$, where  $t_{sw}(t_{f})$ is the function existing by \eqref{eq:tsw}. Moreover, $E(t_f)$ is concave.
\end{theorem}
\begin{proof}
Recall that $\bar{u}$ is constant maximal value of the applied electric power in \eqref{eq:sys}. Also recall from the proof of \Cref{theorem:optcont} that $t_{sw}$ in control (\ref{eq:optcont}) applied to system (\ref{eq:sys}) is uniquely determined with the implicit solution of (\ref{eq:tsw}) for given $t_f$ and fixed parameters $\pA$, $\pB$, $\pC$, $x_0$, and $\bar{u}$. Thus, the energy consumption during an idle period, i.e. idle energy function, can be described as 
\begin{equation} \label{eq:energy-function-form}
E(t_f) = \bar{u} \cdot (t_f - t_{sw}(t_f)).
\end{equation}
Then, for concavity of $E(t_f)$, it remains to show that
\begin{equation} \label{eq:concavity}
    \frac{{\partial^2} E(t_f)}{{\partial} t_f^2} = -\bar u \,  \frac{{\rm d}^{2} t_{sw}}{{\rm d} t_{f}^{2}} 
\end{equation}
is negative $\forall t_{sw}$. Substituting further differentiation of (\ref{implicite}) to (\ref{eq:concavity}) gives
\begin{equation}
\label{implicite2nd}
\frac{{\partial^2} E(t_f)}{{\partial} t_f^2} =-\frac{\pA^{2}\pB x_{0}\,{\rm exp}(\pA t_{sw})}{(\pC x_{0}-\pB\,{\rm exp}(\pA t_{sw}))^{2}} \, \frac{{\rm d} t_{sw}}{{\rm d} t_{f}}.%<0~~\forall t_{sw},
\end{equation}
%
%because $\pB > 0$, $x_0>0$, and $({\rm d}t_{sw})/({\rm d}t_{f})>0$ as it can be seen by \eqref{implicite} since by the assumption \eqref{eq:assmp} $\pB > \pC x_{0} $ and obviously $\pC x_{0} > \pC x_{0} {\rm exp}(-\pA t_{sw}) $ as $\pA > 0, t_{sw} \ge 0$.
%
%As a consequence, $E(t_f)$ is concave and the proof is complete.
%
To prove \eqref{implicite2nd}, first note that by assumption \eqref{eq:assmp}, it holds $\pB > \pC x_{0}$ and therefore, the denominator of the fraction in \eqref{implicite2nd} is positive.
The numerator $\pA^{2}\pB x_{0}\,{\rm exp}(\pA t_{sw})$ is positive as well, since $\pB > 0$ by definition \eqref{eq:sys} and $x_{0} > 0$ by \eqref{eq:constraints}.  Therefore, to prove that \eqref{implicite2nd} is negative, it remains to show that \hbox{$({\rm d}t_{sw})/({\rm d}t_{f})>0\, \forall t_{sw}$}. 
By \eqref{eq:assmp} we have $\pA x_{0} < (\pB-\pC x_0)\bar u$, which also implies $\pA x_{0} < (\pB\,{\rm exp}(\pA t_{sw}) - \pC x_0)\bar u$ since $\pA > 0$ by definition \eqref{eq:sys} and $t_{sw} \geq 0$. It follows that $\frac{\pA x_{0}}{(\pB\,{\rm exp}(\pA t_{sw}) - \pC x_0)\bar u} < 1$, which in turn proves that $({\rm d}t_{sw})/({\rm d}t_{f})>0$.
As a consequence, $E(t_f)$ is concave, and the proof is complete.
\end{proof}
By \Cref{theorem:concavity}, we conclude that problem $\sched$ can be solved in polynomial time for furnaces that can be modelled as (\ref{eq:sys}), and controlled by (\ref{eq:optcont}). In the following section, the proposed approach is shown on a real industrial electric furnace from \v{S}koda Auto.

\section{Case Study: An industrial electric furnace} \label{sec:case-study}

\v{S}koda Auto has a production line employing a ModulTherm\textsuperscript{\textregistered} system by ALD, containing electric vacuum furnaces used for the steel hardening.  
The outer steel shells of the furnaces are cooled by a central cooling system of circulating water at $\sim$\cels{35} to avoid overheating of the system. Thus, we can assume that the ambient temperature ($T_e$) is constant. The operating temperature of the furnaces is set to \cels{960} for the hardening process, which takes about $2.5$ hours on average. 

\begin{figure}
    \centering
    \includegraphics{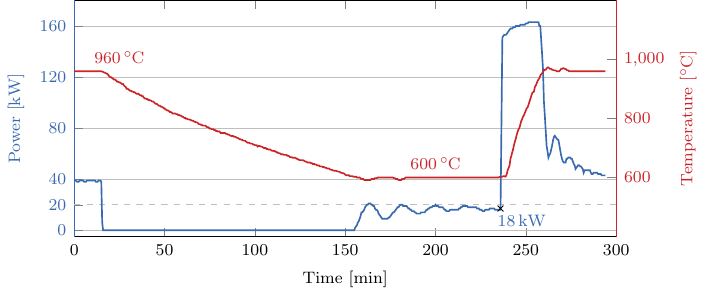}
    \caption{Relationship between the temperature and power when cooling to \cels{600} and heating back to operating temperature}
    \label{fig:data-600}
\end{figure}

The heating of the furnaces has a substantial energy demand across the whole production line.
In a normal regime, all furnaces are turned on and heated to the operating temperature. The operating temperature is preserved even if nothing is being processed. To investigate the potential for energy savings, an experiment has been performed, during which the furnace was cooled to \cels{600}, and its steady-state power consumption was measured. Afterwards, the furnace was heated back to the operating temperature again. Measured data are shown in \Cref{fig:data-600} \citep{thesis:Dusek-2016}. It can be seen that the steady-state power consumption for \cels{600} and \cels{960} is about \kWatt{18} and \kWatt{40}, respectively.

Clearly, if the idle period is long enough, significant energy savings can be achieved by lowering the temperature of the furnace, i.e. turning off the furnace for a longer time and then reheating it back at the right time. This can be achieved by the optimal control law described in the previous section. The rest of the section documents the identification of the furnace in \v{S}koda Auto and shows the resulting idle energy function.

\subsection{Identification of the furnace model}\label{sec:paramestimate} %Parameter estimation of the bilinear model

\begin{figure}[!t]
    \centering
    \includegraphics{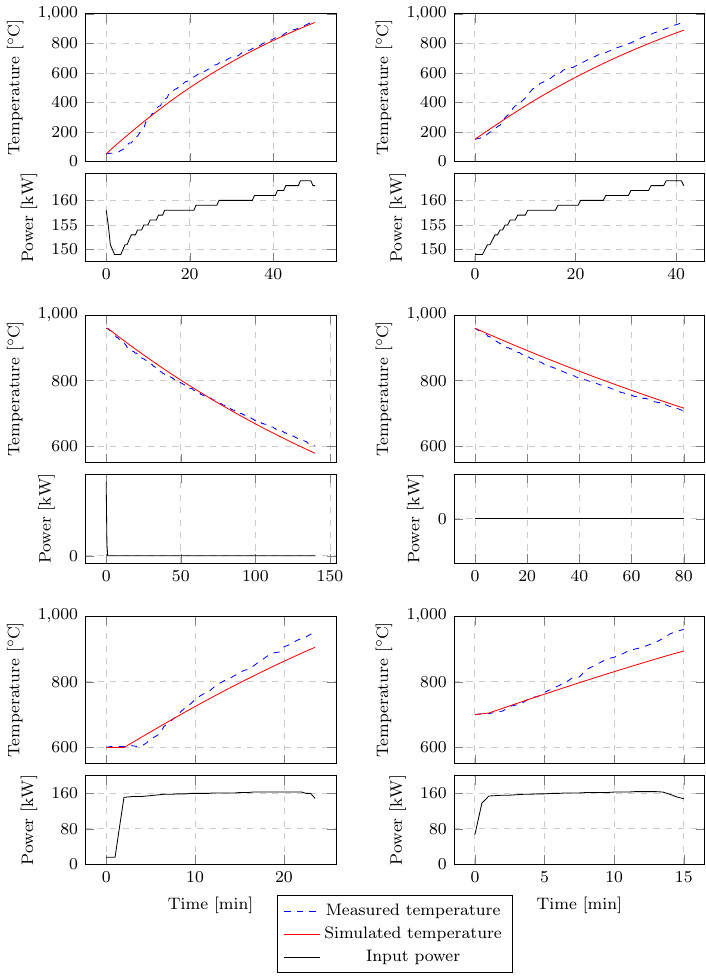}
    \caption{Comparison of the measured data and simulation using a bilinear model.}
    \label{fig:data}
\end{figure}

We employ the bilinear model given by (\ref{eq:sys}) to the furnace mentioned above and estimate the parameters $\pA,\, \pB, \, \pC$ in the model. For this purpose, we use the temperature data collected by \cite{thesis:Dusek-2016} shown by dashed lines in \Cref{fig:data}, with a sampling time of \secs{30}. 
The system parameters are estimated as
\begin{equation} \label{eq:sysparam}
\pA = 0.003821964, \quad \pB = 0.175187494, \quad \pC = 0.000094367
\end{equation}
by the least-squares method using the measured temperature samples and their derivatives obtained via a polynomial regression. The simulated response of the system (\ref{eq:sys}) with (\ref{eq:sysparam}) is illustrated by red lines in \Cref{fig:data}, when the experimental input power is applied. It is seen that the utilised bilinear model provides a reasonable fit to the measured temperature values of the furnace. Note, that all the measurements were carried out during production and it was not possible to test arbitrary input signals (i.e. power). 
Nevertheless, the mean absolute percentage error over all experiments for the identified model is found as \percents{4.49}, which is sufficiently accurate for the system identification.

\subsection{Idle energy function of the furnace} \label{sec:skoda-energy-func} 
\begin{figure}
    \centering
    \includegraphics{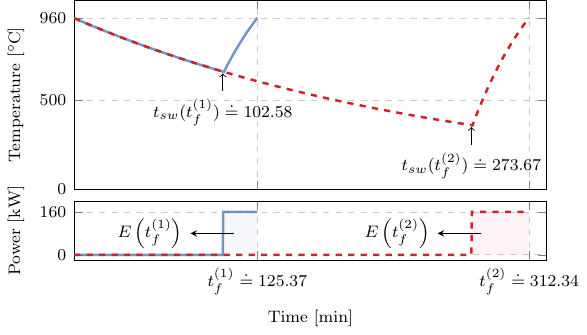}
    \caption{Example of the optimal control for two different terminal times $t_f^{(1)}$ and $t_{f}^{(2)}$.}
    \label{fig:optcont}
\end{figure}

To reveal the idle energy function of the furnace, let us first demonstrate the furnace temperature response under the proposed energy-optimal control law given by \Cref{theorem:optcont}.
In \Cref{fig:optcont}, the time response of the furnace model (\ref{eq:sys}) with the parameters (\ref{eq:sysparam}) is illustrated via simulations for two different terminal times ($t_f^{(1)}$ and $t_f^{(2)}$), i.e. idle periods, when the optimal control (\ref{eq:optcont}) is applied. Indeed, the applied input power is switched from zero to the maximum applicable power $\bar{u}$ (160 kW) at the appropriate switching times $t_{sw}(t_f^{(1)})$ and $t_{sw}(t_f^{(2)})$ calculated by \eqref{eq:tsw}, to ensure reaching the operating temperature (\cels{960}) at the end of each idle period. 
The corresponding minimal energy consumptions $E(t_f^{(1)})$ and $E(t_f^{(2)})$ (calculated by \eqref{eq:energy-function-form}) are also illustrated in the lower part of \Cref{fig:optcont}.

\begin{figure}
    \centering
    \includegraphics{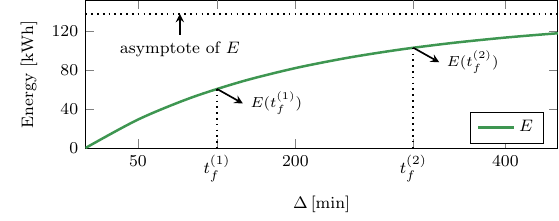}
    \caption{Idle energy function $\fEnoarg$ and two different idle period lengths $t_f^{(1)}$ and $t_f^{(2)}$ with the corresponding idle energy consumption $\fE{t_f^{(1)}}$ and $\fE{t_f^{(2)}}$.}
    \label{fig:enery-graphs}
\end{figure}

Performing the above explained calculations for an appropriate sampling of the idle period length $t_{f}$, one can obtain the idle energy function $\fEnoarg$, as shown in \Cref{fig:enery-graphs}. 
Function $\fEnoarg$ is bounded by a constant shown by the dashed line, which is the energy for heating the machine from the ambient temperature (\cels{35}) to the operating temperature. Clearly, it is seen that $\fEnoarg$ is concave, as declared by \Cref{theorem:concavity}.
\begin{remark} \label{rem:application}
Note that for the real furnace application the proposed control may not be precisely optimal, and the operating temperature may not be reached exactly at $t=t_f$, inherently due to the uncertain dynamics and the approximate modelling. Nevertheless, the proposed approximation is acceptable for achieving almost optimal control in practice. The reach of the operating temperature can be guaranteed with a simple \emph{if case} control as is actually done in switching to feedback control around the operating point in practical process control approaches.
\end{remark}


\section{Comparison to the state-of-the-art approaches} \label{sec:experiments}
\begin{figure}
    \centering
    \begin{tabular}{ccc}
        \includegraphics{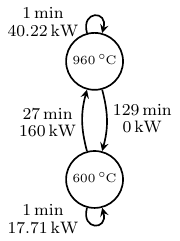} &
        \includegraphics{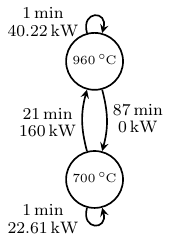} &
        \includegraphics{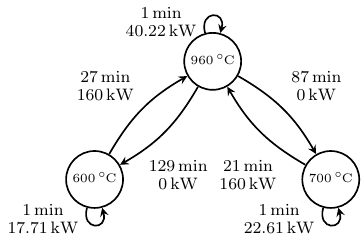} \\
        a) $G_{600}$ & b) $G_{700}$ & c) $G_{600,700}$ \\
    \end{tabular}
    
    \caption{Examples of the transition graphs for the furnace model \eqref{eq:sys} with parameters \eqref{eq:sysparam}.}
    \label{fig:sota-trans-graphs}
\end{figure}

As it was explained in the introduction, conventional scheduling approaches to idle energy optimisation assume only a small number of machine modes to approximate the dynamics of the machine \citep{2007:Mouzon,2014:Shrouf,2017:Che-unrelated,2019:abikarram}. To represent the machine modes, the authors typically use the static transition graph, where the vertices represent the modes, and the edges represent the available transitions between them. The edges are labelled by the time, which is needed for the transition, and the power, which is consumed during the transition. Examples of the transition graphs for the furnace model \eqref{eq:sys} with parameters \eqref{eq:sysparam} are shown in \Cref{fig:sota-trans-graphs}. These graphs represent simple scenarios, with a single processing mode (\cels{960}) and one ($G_{600}$, $G_{700}$), or two ($G_{600,700}$), standby modes. The standby modes correspond to allowed temperatures, to which the furnace can be cooled during the idle periods (here \cels{600}, and \cels{700}).

The primal aim of this section is to show, why representation via an idle energy function is better than a transition graph. This is illustrated by an experiment described in Section~\ref{sec:exp-1}. Secondly, we compare complexity of the algorithm for problem $\sched$ described in Section~\ref{sec:complexity-analysis} with the state-of-the-art approaches. This analysis is described in Section~\ref{sec:experiment-sota-approach}.

\subsection{Benchmark instances} \label{sec:instanges-generation}

Considering the behaviour of the machine, we use the idle energy function $\fEnoarg$ depicted in \Cref{fig:enery-graphs} for the minimisation of the objective \eqref{eq:objective}. Our approach is compared to the dynamic programming adopted from \cite{2019:Aghelinejad}, which represents the behaviour of the machine by a finite transition graph. For the comparison, we use the transition graphs $G_{600}$, $G_{700}$, and $G_{600,700}$ depicted in \Cref{fig:sota-trans-graphs}.

Now we describe, how we generate the tasks parameters for the benchmarks instances.
A set of 6750 instances was generated using  \Cref{alg:tasks}. Specifically, 10 instances were generated for each combination of $\cNtasks \in \{30, 40, 50\}$, $\paramAlpha \in \{0.2, 0.4,\dots, 3.0\}$, and $\paramBeta \in \{0.2, 0.4, \dots, 3.0\}$. A wide range of parameters $\paramAlpha$ and $\paramBeta$ was used to generate data of different characteristics. Constants $\cPmin$ and $\cPmax$, denoting the minimal and the maximal processing time, were set to $1$ and $300$, respectively. Note that \Cref{alg:tasks} is designed such that only feasible instances are generated. By $\fUniform{a}{b}$, we denote integer uniform distribution on set $\{a, a+1, \dots, b\}$; here $\fExponential{x}$ denotes exponential distribution with scale parameter $x$.
\begin{algorithm}[!h]
\SetKwInOut{Input}{input}
\SetKwInOut{Output}{output}
\SetKwFunction{Avg}{Average}

\Input{Number of tasks $\cNtasks$, bounds on processing time $\cPmin$, $\cPmax$, parameters $\paramAlpha$, $\paramBeta$
}
\Output{Vectors $\vecR$, $\vecD$, $\vecP$}

\BlankLine

\tcp{generate processing times}
\lForEach{$i \gets 1$ \KwTo $\cNtasks$}{$\cP{\idxTask} \sim \fUniform{\cPmin}{\cPmax}$} 

\BlankLine

\tcp{generate release times and deadlines}

$\cR{1} := 0$ \;
$\cD{1} \sim \lceil \cR{1} + \cP{1} + \fExponential{\paramBeta \cdot \Avg(\vecP)} \rceil$ \;
\ForEach{$i \gets 2$ \KwTo $\cNtasks$}{
$\cR{\idxTask} \sim \lceil \cR{\idxTask-1} + \cP{\idxTask-1} + \fExponential{\paramAlpha \cdot \Avg(\vecP)} \rceil $\;
$\cD{\idxTask} \sim \lceil \cR{\idxTask} + \cP{\idxTask} + \fExponential{\paramBeta \cdot \Avg(\vecP)} \rceil$ \;
} 

\BlankLine

\tcp{propagate deadlines by \eqref{eq:propagate-deadlines} (release times are already propagated)}
\lForEach{$i \gets (\cNtasks-1)$ \KwTo 1}{$\cD{\idxTask} := \min \{\cD{\idxTask+1} - \cP{\idxTask+1}, \cD{\idxTask}\}$}

 \caption{Generation of task parameters} \label{alg:tasks}
\end{algorithm}

One of the factors influencing the final energy savings is the utilisation of the machine, which is calculated as the ratio between the sum of processing times and length of the scheduling horizon, i.e. ${\sum_{i=1}^{\cNtasks} \cP{i}}\, / \, (\cD{\cNtasks} - \cR{1})$. Based on the machine utilisation, the generated instances were divided, as indicated by \Cref{tab:exp-instances-count}. 

\begin{table}
\tbl{Number of generated instances with respect to utilisation (columns) and number of tasks (rows).}
{
    \bgroup
    \def\arraystretch{1.2}
    \begin{tabular}{rrrrrrrrrr}
    \toprule
    & & \multicolumn{8}{c}{Utilisation} \\ \cmidrule{3-10}
    & & (0.1, 0.2] & (0.2, 0.3] & (0.3, 0.4] & (0.4, 0.5] & (0.5, 0.6] & (0.6, 0.7] & (0.7, 0.8] & (0.8, 0.9] \\ 
    \midrule
    & 30 &  12 & 532 & 621 & 391 & 286 & 193 & 125 & 90  \\ 
    $\cNtasks$ & 40 &  2 & 508 & 672 & 383 & 273 & 191 & 113 & 108  \\ 
    & 50 &  5 & 520 & 672 & 376 & 252 & 195 & 121 & 109  \\ 
    \midrule
    \multicolumn{2}{r}{Total} & 19 & 1560 & 1965 & 1150 & 811 & 579 & 359 & 307  \\ 
    \bottomrule
    \end{tabular}      
    \egroup
}
 \label{tab:exp-instances-count}
\end{table}

\subsection{Transition graph vs. idle energy functions} \label{sec:exp-1}
For the experiment, we optimised all generated instances with respect to the idle energy functions $\fEnoarg$ (our approach), and transition graphs $G_{600}$, $G_{700}$, and $G_{600,700}$ (representing the state-of-the-art approaches assuming only a small number of modes). The instances with transition graphs $G_{600}$, $G_{700}$, and $G_{600,700}$ were optimised using the dynamic programming adopted from \citep{2019:Aghelinejad}.

To compare the results, we define the average power per idle time $\overline{P}$ as
\begin{equation}
    \overline{P} = \frac{\fEtotalNoarg^{\star}}{(\cD{\cNtasks} - \cR{1}) - \sum\limits_{\idxTask = 1}^{\cNtasks} \cP{\idxTask}},
\end{equation}
where $\fEtotalNoarg^{\star}$ is the optimal total idle energy consumption (with respect to given idle energy function or transition graph). It is assumed that the machine is underutilised, i.e. $(\cD{\cNtasks} - \cR{1}) - \sum_{\idxTask = 1}^{\cNtasks} \cP{\idxTask} > 0$. For the considered models, it holds that $0 \leq \overline{P} \leq \overline{P}_{\text{max}}$, where $\overline{P}_{\text{max}}$ is the theoretical worst case, representing the situation when the furnace is heated to the operating temperature all the time.

\begin{figure*}
    \centering
    \includegraphics{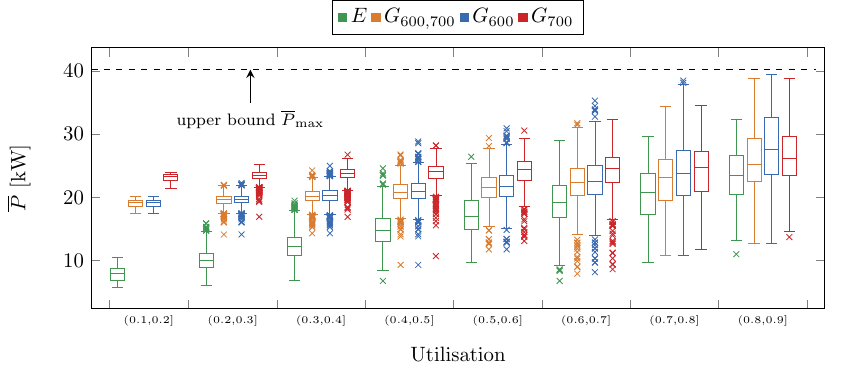}
    \caption{Average power per idle time $\overline{P}$ depending on the modelling of the machine dynamics and utilisation of the machine.}
    \label{fig:boxplot-comp}
\end{figure*}

Results for different utilisations of the machines are shown in the form of boxplots in \Cref{fig:boxplot-comp}. Clearly, our approach using $\fEnoarg$ dominates all the transition graphs, as the power saving modes modelled by $G_{600}$, $G_{700}$, and $G_{600,700}$ are only a subset of all possible modes implicitly encoded in $\fEnoarg$. The difference increases when utilisation is lowered as the idle periods become longer. For example, the average $\overline{P}$ for $\fEnoarg$ is less than half compared to $G_{600,700}$ for utilisation $(0.1,0.2]$.

It can be seen that $\overline{P}$ optimised with respect to $G_{600}$ nearly converges to steady-state power compensating for the energy loss at \cels{600}, which is approximately \kWatt{18}. Similar observation also holds for $G_{700}$, and $G_{600,700}$.
Using $G_{700}$ is slightly better than $G_{600}$ only when the utilisation is high because shorter idle periods do not allow the standby mode corresponding to \cels{600} to be reached.

\subsection{Time complexity comparison} \label{sec:experiment-sota-approach}

The authors of conventional scheduling approaches to idle energy optimisation use the ILP formalism for the modelling \citep{2007:Mouzon,2014:Shrouf,2017:Che-unrelated,2019:abikarram}. 
The scheduling horizon is discretised into a set of intervals $H$ (e.g. one minute long), and for each interval $\idxInterval \in H$ and each possible mode of the machine $m$, binary variables encode whether the machine operates in mode $m$ during interval $\idxInterval$ or not~\citep{2019:abikarram, 2014:Shrouf, 2018:Aghelinejad}.
The main weakness in these approaches is that the size of the model depends on the number of intervals in $H$ as well as on the number of machine states. Therefore, the model can be used successfully only for small instances of the problem. When long scheduling horizon is considered (e.g. 7200 minutes in a work-week), building and optimisation of such model become intractable.

To the best of our knowledge, the nearest polynomial-time approach that can be adopted to solve the problem addressed in this paper is described in \cite{2019:Aghelinejad}. Assuming that the scheduling horizon is discretised and the order of the tasks if fixed, the problem can be transformed to the shortest path problem.
Aghelinejad et al. construct graph $G$ having $|H|$ layers, each of which is containing about $\sum_{\idxTask \in \setTask} \cP{\idxTask}$ nodes. Node $n(\idxTask,\idxInterval)$ in layer $\idxInterval$ encodes that $\idxTask$ intervals were spent for the processing from the beginning till time $\idxInterval$. The graph contains $\mathcal{O}(|H| \sum_{\idxTask \in \setTask} \cP{\idxTask})$ nodes, and $\mathcal{O}(|H|^2 \sum_{\idxTask \in \setTask} \cP{\idxTask})$ edges.
The shortest path representing the schedule with lowest energy consumption can be found by dynamic programming in $\mathcal{O}(|H|^2 \sum_{\idxTask \in \setTask} \cP{\idxTask})$.
In the original paper \citep{2019:Aghelinejad}, the authors did not assume release times and deadlines. However, their approach can be easily extended by removing the edges, which would cause the processing of the task $\idxTask$ outside of its execution window defined by $[\cR{\idxTask}, \cD{\idxTask}]$.
Further, in the case of the problem studied in this paper, it is not necessary to model every unit of tasks' processing times. Thus, term $\sum_{\idxTask \in \setTask} \cP{\idxTask}$ can be substituted by $n$ (processing units corresponding to a single job can be joined together). Therefore, the complexity of solving our problem by the approach described in \cite{2019:Aghelinejad} is $\mathcal{O}(|H|^2 \cNtasks)$ assuming that the scheduling horizon is discretised into $|H|$ intervals.

In comparison, the energy graph proposed in this paper contains $\mathcal{O}(\cNtasks)$ nodes and $\mathcal{O}(\cNtasks^2)$ edges and can be constructed in $\mathcal{O}(\cNtasks^3)$ steps. The overall complexity of our approach is therefore $\mathcal{O}(\cNtasks^3)$. Taking into account that for a real production $|H|$ is typically larger than $\cNtasks$, the complexity of our approach is significantly better.

Summarising, we believe that there are two main drawbacks in the adaptation of the state-of-the-art approaches (including both the ILP models as well as the graph proposed in \cite{2019:Aghelinejad}). First, the complexity of the state-of-the-art approaches sharply grows with the length of the scheduling horizon $H$, while our approach is independent on it. Second, a finite number of machine modes cannot fully describe the behaviour of more complex systems. For example, see function $\fEnoarg$ in \Cref{fig:enery-graphs} representing the energy consumption w.r.t. the length of the idle period for our case study. The shape of this function cannot be reasonably approximated by a simple transition graph with several modes only.

\section{Conclusions} \label{sec:conclusions}

This paper has two aims. The first one is to show that for some machines, e.g. furnaces and other heat-intensive systems, when approximating their dynamics by a simple transition graph, the scheduling algorithm cannot achieve the maximum energy savings.
For such systems, we propose a different concept incorporating the complete dynamics and the optimal control of the machine into the idle energy function, which represents the energy consumption of the machine much better.
The analysis in \Cref{sec:exp-1} on an electric furnace from \v{S}koda Auto company shows the significant difference between these two concepts.
Second, we show that problem $\sched$ can be solved in polynomial time, assuming that the idle energy function is concave. The time complexity of our algorithm is better than the complexity of related state-of-the-art algorithms, as it is explained in \Cref{sec:experiment-sota-approach}.

Our analysis is focused on heat-intensive processes, as the most typical applications in the domain of idle energy optimisation and scheduling. Indeed, our analysis cannot be applied to an arbitrary machine, and we cannot analyse every possible one. Nevertheless, many energy demanding systems have very similar properties, often resulting in a concave idle energy function. Moreover, the concept of energy function allows integrating the system dynamics and its energy-optimal control, studied in the control engineering domain, into the scheduling domain. As we believe, this synergy is essential for achieving maximal energetic efficiency. A related example can be found in papers \cite{2019:Bukata, 2018:Bukata} studying energy optimisation of robotic cells, where very complex dynamics of a robotic manipulator is also encoded into an energy function. Those papers do not study idle energy consumption but address the relation between the speed limit of a robot movement and its energy consumption. Unlike the case with the furnaces, this function is convex; nevertheless, the idea of the decomposition is the same. Therefore, as we believe, there are other applications where the complex dynamics of a machine can be expressed using a nonlinear function and exploited in a scheduling algorithm to achieve the best savings. Therefore, finding other scenarios where an energy function can be used is the real challenge for future research.

\section*{Acknowledgement}

The authors want to thank Josef Du\v{s}ek from \v{S}koda Auto for carrying out the analysis of the hardening line. Furthermore, we are glad to acknowledge the help of Jan Sko\v{c}ilas from the Faculty of Mechanical Engineering, Czech Technical University in Prague, who provided insights into the field of electric furnaces. Last but not least, we thank our colleague Mat\v{e}j Novotn\'{y}, for his ideas that helped us analyse the idle energy function.

This work was funded by Ministry of Education, Youth and Sport of the Czech Republic within the project Cluster 4.0 number CZ.02.1.01/0.0/0.0/16\_026/0008432; and European Regional Development Fund under the project Robotics for Industry 4.0 (reg. no. CZ.02.1.01/0.0/0.0/15\_0030000470). 

\bibliographystyle{plainnat}
\bibliography{refs}

\end{document}